\documentclass[longauth]{aa}

\usepackage{graphicx}
\usepackage{natbib}

\usepackage{txfonts}
\usepackage{hyperref}
\hypersetup{
    colorlinks=true,
    linkcolor=blue,
    filecolor=magenta,      
    urlcolor=blue,
    citecolor=blue
}

\makeatletter
\renewcommand*\aa@pageof{, page \thepage{} of \pageref*{LastPage}}
\makeatother

\usepackage[switch, modulo]{lineno}

\usepackage{euclid}
\usepackage{orcidlink}
\let\orcid\orcidlink

\usepackage[nolist,nohyperlinks]{acronym}

\usepackage[capitalise]{cleveref}
\crefname{section}{Sect.}{Sects.}

\begin{document}

\acrodef{CFHT}{Canada-France-Hawaii Telescope}
\acrodef{DES}{Dark Energy Survey}
\acrodef{EDS}{Euclid Deep Survey}
\acrodef{ERO}{Early Release Observations}
\acrodef{EWS}{Euclid Wide Survey}
\acrodef{PSF}{point spread function}
\acrodef{FWHM}{full width at half maximum}
\acrodef{NISP}{Near-Infrared Spectrometer and Photometer}
\acrodef{VIS}{visible imager}
\newacro{SLED}{Strong Lens Database}
\newacro{CSST}{Chinese Space Station Telescope} 
\newcommand{\prob}[2]{P(#1\vert#2)}
\newcommand{\prior}[1]{P(#1)}

\author{J.~A.~Acevedo~Barroso\orcid{0000-0002-9654-1711}\thanks{\email{javier.acevedobarroso@epfl.ch}}\inst{\ref{aff1}}
\and C.~M.~O'Riordan\orcid{0000-0003-2227-1998}\inst{\ref{aff2}}
\and B.~Cl\'ement\orcid{0000-0002-7966-3661}\inst{\ref{aff1},\ref{aff3}}
\and C.~Tortora\orcid{0000-0001-7958-6531}\inst{\ref{aff4}}
\and T.~E.~Collett\orcid{0000-0001-5564-3140}\inst{\ref{aff5}}
\and F.~Courbin\orcid{0000-0003-0758-6510}\inst{\ref{aff1},\ref{aff6},\ref{aff7}}
\and R.~Gavazzi\orcid{0000-0002-5540-6935}\inst{\ref{aff8},\ref{aff9}}
\and R.~B.~Metcalf\orcid{0000-0003-3167-2574}\inst{\ref{aff10},\ref{aff11}}
\and V.~Busillo\orcid{0009-0000-6049-1073}\inst{\ref{aff4},\ref{aff12},\ref{aff13}}
\and I.~T.~Andika\orcid{0000-0001-6102-9526}\inst{\ref{aff14},\ref{aff2}}
\and R.~Cabanac\orcid{0000-0001-6679-2600}\inst{\ref{aff15}}
\and H.~M.~Courtois\orcid{0000-0003-0509-1776}\inst{\ref{aff16}}
\and J.~Crook-Mansour\orcid{0000-0001-7466-1192}\inst{\ref{aff17}}
\and L.~Delchambre\orcid{0000-0003-2559-408X}\inst{\ref{aff18}}
\and G.~Despali\orcid{0000-0001-6150-4112}\inst{\ref{aff10},\ref{aff11},\ref{aff19}}
\and L.~R.~Ecker\inst{\ref{aff20},\ref{aff21}}
\and A.~Franco\orcid{0000-0002-4761-366X}\inst{\ref{aff22},\ref{aff23},\ref{aff24}}
\and P.~Holloway\orcid{0009-0002-8896-6100}\inst{\ref{aff25}}
\and N.~Jackson\inst{\ref{aff26}}
\and K.~Jahnke\orcid{0000-0003-3804-2137}\inst{\ref{aff27}}
\and G.~Mahler\orcid{0000-0003-3266-2001}\inst{\ref{aff18},\ref{aff28},\ref{aff29}}
\and L.~Marchetti\orcid{0000-0003-3948-7621}\inst{\ref{aff17},\ref{aff30}}
\and P.~Matavulj\orcid{0000-0003-0229-7189}\inst{\ref{aff31}}
\and A.~Melo\orcid{0000-0002-6449-3970}\inst{\ref{aff2},\ref{aff14}}
\and M.~Meneghetti\orcid{0000-0003-1225-7084}\inst{\ref{aff11},\ref{aff19}}
\and L.~A.~Moustakas\orcid{0000-0003-3030-2360}\inst{\ref{aff32}}
\and O.~M\"uller\orcid{0000-0003-4552-9808}\inst{\ref{aff1}}
\and A.~A.~Nucita\inst{\ref{aff23},\ref{aff22},\ref{aff24}}
\and A.~Paulino-Afonso\orcid{0000-0002-0943-0694}\inst{\ref{aff33},\ref{aff34}}
\and J.~Pearson\orcid{0000-0001-8555-8561}\inst{\ref{aff35}}
\and K.~Rojas\orcid{0000-0003-1391-6854}\inst{\ref{aff5}}
\and C.~Scarlata\orcid{0000-0002-9136-8876}\inst{\ref{aff36}}
\and S.~Schuldt\orcid{0000-0003-2497-6334}\inst{\ref{aff37},\ref{aff38}}
\and S.~Serjeant\orcid{0000-0002-0517-7943}\inst{\ref{aff35}}
\and D.~Sluse\orcid{0000-0001-6116-2095}\inst{\ref{aff18}}
\and S.~H.~Suyu\orcid{0000-0001-5568-6052}\inst{\ref{aff14},\ref{aff2}}
\and M.~Vaccari\orcid{0000-0002-6748-0577}\inst{\ref{aff17},\ref{aff39},\ref{aff30}}
\and A.~Verma\orcid{0000-0002-0730-0781}\inst{\ref{aff25}}
\and G.~Vernardos\orcid{0000-0001-8554-7248}\inst{\ref{aff40}}
\and M.~Walmsley\orcid{0000-0002-6408-4181}\inst{\ref{aff41},\ref{aff26}}
\and H.~Bouy\orcid{0000-0002-7084-487X}\inst{\ref{aff42},\ref{aff43}}
\and G.~L.~Walth\orcid{0000-0002-6313-6808}\inst{\ref{aff44}}
\and D.~M.~Powell\orcid{0000-0002-4912-9943}\inst{\ref{aff2}}
\and M.~Bolzonella\orcid{0000-0003-3278-4607}\inst{\ref{aff11}}
\and J.-C.~Cuillandre\orcid{0000-0002-3263-8645}\inst{\ref{aff45}}
\and M.~Kluge\orcid{0000-0002-9618-2552}\inst{\ref{aff21}}
\and T.~Saifollahi\orcid{0000-0002-9554-7660}\inst{\ref{aff46}}
\and M.~Schirmer\orcid{0000-0003-2568-9994}\inst{\ref{aff27}}
\and C.~Stone\orcid{0000-0002-9086-6398}\inst{\ref{aff47}}
\and A.~Acebron\orcid{0000-0003-3108-9039}\inst{\ref{aff48}}
\and L.~Bazzanini\orcid{0000-0003-0727-0137}\inst{\ref{aff49},\ref{aff11}}
\and A.~D\'iaz-S\'anchez\orcid{0000-0003-0748-4768}\inst{\ref{aff50}}
\and N.~B.~Hogg\orcid{0000-0001-9346-4477}\inst{\ref{aff51}}
\and L.~V.~E.~Koopmans\orcid{0000-0003-1840-0312}\inst{\ref{aff52}}
\and S.~Kruk\orcid{0000-0001-8010-8879}\inst{\ref{aff53}}
\and L.~Leuzzi\inst{\ref{aff10},\ref{aff11}}
\and A.~Manj\'on-Garc\'ia\orcid{0000-0002-7413-8825}\inst{\ref{aff50}}
\and F.~Mannucci\orcid{0000-0002-4803-2381}\inst{\ref{aff54}}
\and B.~C.~Nagam\orcid{0000-0002-3724-7694}\inst{\ref{aff52}}
\and R.~Pearce-Casey\inst{\ref{aff35}}
\and L.~Scharr\'e\orcid{0000-0003-2551-4430}\inst{\ref{aff55}}
\and J.~Wilde\orcid{0000-0002-4460-7379}\inst{\ref{aff35}}
\and B.~Altieri\orcid{0000-0003-3936-0284}\inst{\ref{aff53}}
\and A.~Amara\inst{\ref{aff56}}
\and S.~Andreon\orcid{0000-0002-2041-8784}\inst{\ref{aff57}}
\and N.~Auricchio\orcid{0000-0003-4444-8651}\inst{\ref{aff11}}
\and C.~Baccigalupi\orcid{0000-0002-8211-1630}\inst{\ref{aff58},\ref{aff59},\ref{aff60},\ref{aff61}}
\and M.~Baldi\orcid{0000-0003-4145-1943}\inst{\ref{aff62},\ref{aff11},\ref{aff19}}
\and A.~Balestra\orcid{0000-0002-6967-261X}\inst{\ref{aff63}}
\and S.~Bardelli\orcid{0000-0002-8900-0298}\inst{\ref{aff11}}
\and A.~Basset\inst{\ref{aff64}}
\and P.~Battaglia\orcid{0000-0002-7337-5909}\inst{\ref{aff11}}
\and R.~Bender\orcid{0000-0001-7179-0626}\inst{\ref{aff21},\ref{aff20}}
\and D.~Bonino\orcid{0000-0002-3336-9977}\inst{\ref{aff65}}
\and E.~Branchini\orcid{0000-0002-0808-6908}\inst{\ref{aff66},\ref{aff67},\ref{aff57}}
\and M.~Brescia\orcid{0000-0001-9506-5680}\inst{\ref{aff12},\ref{aff4},\ref{aff13}}
\and J.~Brinchmann\orcid{0000-0003-4359-8797}\inst{\ref{aff34},\ref{aff68}}
\and A.~Caillat\inst{\ref{aff8}}
\and S.~Camera\orcid{0000-0003-3399-3574}\inst{\ref{aff69},\ref{aff70},\ref{aff65}}
\and G.~P.~Candini\orcid{0000-0001-9481-8206}\inst{\ref{aff71}}
\and V.~Capobianco\orcid{0000-0002-3309-7692}\inst{\ref{aff65}}
\and C.~Carbone\orcid{0000-0003-0125-3563}\inst{\ref{aff38}}
\and J.~Carretero\orcid{0000-0002-3130-0204}\inst{\ref{aff72},\ref{aff73}}
\and S.~Casas\orcid{0000-0002-4751-5138}\inst{\ref{aff74}}
\and M.~Castellano\orcid{0000-0001-9875-8263}\inst{\ref{aff75}}
\and G.~Castignani\orcid{0000-0001-6831-0687}\inst{\ref{aff11}}
\and S.~Cavuoti\orcid{0000-0002-3787-4196}\inst{\ref{aff4},\ref{aff13}}
\and A.~Cimatti\inst{\ref{aff76}}
\and C.~Colodro-Conde\inst{\ref{aff77}}
\and G.~Congedo\orcid{0000-0003-2508-0046}\inst{\ref{aff78}}
\and C.~J.~Conselice\orcid{0000-0003-1949-7638}\inst{\ref{aff26}}
\and L.~Conversi\orcid{0000-0002-6710-8476}\inst{\ref{aff79},\ref{aff53}}
\and Y.~Copin\orcid{0000-0002-5317-7518}\inst{\ref{aff80}}
\and L.~Corcione\orcid{0000-0002-6497-5881}\inst{\ref{aff65}}
\and M.~Cropper\orcid{0000-0003-4571-9468}\inst{\ref{aff71}}
\and A.~Da~Silva\orcid{0000-0002-6385-1609}\inst{\ref{aff81},\ref{aff82}}
\and H.~Degaudenzi\orcid{0000-0002-5887-6799}\inst{\ref{aff83}}
\and G.~De~Lucia\orcid{0000-0002-6220-9104}\inst{\ref{aff59}}
\and J.~Dinis\orcid{0000-0001-5075-1601}\inst{\ref{aff81},\ref{aff82}}
\and F.~Dubath\orcid{0000-0002-6533-2810}\inst{\ref{aff83}}
\and X.~Dupac\inst{\ref{aff53}}
\and S.~Dusini\orcid{0000-0002-1128-0664}\inst{\ref{aff84}}
\and M.~Farina\orcid{0000-0002-3089-7846}\inst{\ref{aff85}}
\and S.~Farrens\orcid{0000-0002-9594-9387}\inst{\ref{aff45}}
\and S.~Ferriol\inst{\ref{aff80}}
\and M.~Frailis\orcid{0000-0002-7400-2135}\inst{\ref{aff59}}
\and E.~Franceschi\orcid{0000-0002-0585-6591}\inst{\ref{aff11}}
\and S.~Galeotta\orcid{0000-0002-3748-5115}\inst{\ref{aff59}}
\and B.~Garilli\orcid{0000-0001-7455-8750}\thanks{Deceased}\inst{\ref{aff38}}
\and K.~George\orcid{0000-0002-1734-8455}\inst{\ref{aff20}}
\and W.~Gillard\orcid{0000-0003-4744-9748}\inst{\ref{aff86}}
\and B.~Gillis\orcid{0000-0002-4478-1270}\inst{\ref{aff78}}
\and C.~Giocoli\orcid{0000-0002-9590-7961}\inst{\ref{aff11},\ref{aff87}}
\and P.~G\'omez-Alvarez\orcid{0000-0002-8594-5358}\inst{\ref{aff88},\ref{aff53}}
\and A.~Grazian\orcid{0000-0002-5688-0663}\inst{\ref{aff63}}
\and F.~Grupp\inst{\ref{aff21},\ref{aff20}}
\and L.~Guzzo\orcid{0000-0001-8264-5192}\inst{\ref{aff37},\ref{aff57}}
\and S.~V.~H.~Haugan\orcid{0000-0001-9648-7260}\inst{\ref{aff89}}
\and H.~Hoekstra\orcid{0000-0002-0641-3231}\inst{\ref{aff90}}
\and W.~Holmes\inst{\ref{aff32}}
\and I.~Hook\orcid{0000-0002-2960-978X}\inst{\ref{aff91}}
\and F.~Hormuth\inst{\ref{aff92}}
\and A.~Hornstrup\orcid{0000-0002-3363-0936}\inst{\ref{aff93},\ref{aff94}}
\and M.~Jhabvala\inst{\ref{aff95}}
\and B.~Joachimi\orcid{0000-0001-7494-1303}\inst{\ref{aff96}}
\and E.~Keih\"anen\orcid{0000-0003-1804-7715}\inst{\ref{aff97}}
\and S.~Kermiche\orcid{0000-0002-0302-5735}\inst{\ref{aff86}}
\and A.~Kiessling\orcid{0000-0002-2590-1273}\inst{\ref{aff32}}
\and B.~Kubik\orcid{0009-0006-5823-4880}\inst{\ref{aff80}}
\and M.~Kunz\orcid{0000-0002-3052-7394}\inst{\ref{aff98}}
\and H.~Kurki-Suonio\orcid{0000-0002-4618-3063}\inst{\ref{aff99},\ref{aff100}}
\and D.~Le~Mignant\orcid{0000-0002-5339-5515}\inst{\ref{aff8}}
\and S.~Ligori\orcid{0000-0003-4172-4606}\inst{\ref{aff65}}
\and P.~B.~Lilje\orcid{0000-0003-4324-7794}\inst{\ref{aff89}}
\and V.~Lindholm\orcid{0000-0003-2317-5471}\inst{\ref{aff99},\ref{aff100}}
\and I.~Lloro\inst{\ref{aff101}}
\and G.~Mainetti\orcid{0000-0003-2384-2377}\inst{\ref{aff102}}
\and E.~Maiorano\orcid{0000-0003-2593-4355}\inst{\ref{aff11}}
\and O.~Mansutti\orcid{0000-0001-5758-4658}\inst{\ref{aff59}}
\and S.~Marcin\inst{\ref{aff31}}
\and O.~Marggraf\orcid{0000-0001-7242-3852}\inst{\ref{aff103}}
\and M.~Martinelli\orcid{0000-0002-6943-7732}\inst{\ref{aff75},\ref{aff104}}
\and N.~Martinet\orcid{0000-0003-2786-7790}\inst{\ref{aff8}}
\and F.~Marulli\orcid{0000-0002-8850-0303}\inst{\ref{aff10},\ref{aff11},\ref{aff19}}
\and R.~Massey\orcid{0000-0002-6085-3780}\inst{\ref{aff29}}
\and E.~Medinaceli\orcid{0000-0002-4040-7783}\inst{\ref{aff11}}
\and M.~Melchior\inst{\ref{aff31}}
\and Y.~Mellier\inst{\ref{aff105},\ref{aff9}}
\and E.~Merlin\orcid{0000-0001-6870-8900}\inst{\ref{aff75}}
\and G.~Meylan\inst{\ref{aff1}}
\and M.~Moresco\orcid{0000-0002-7616-7136}\inst{\ref{aff10},\ref{aff11}}
\and L.~Moscardini\orcid{0000-0002-3473-6716}\inst{\ref{aff10},\ref{aff11},\ref{aff19}}
\and E.~Munari\orcid{0000-0002-1751-5946}\inst{\ref{aff59},\ref{aff58}}
\and R.~Nakajima\inst{\ref{aff103}}
\and C.~Neissner\orcid{0000-0001-8524-4968}\inst{\ref{aff106},\ref{aff73}}
\and R.~C.~Nichol\orcid{0000-0003-0939-6518}\inst{\ref{aff56}}
\and S.-M.~Niemi\inst{\ref{aff107}}
\and J.~W.~Nightingale\orcid{0000-0002-8987-7401}\inst{\ref{aff108},\ref{aff29}}
\and C.~Padilla\orcid{0000-0001-7951-0166}\inst{\ref{aff106}}
\and S.~Paltani\orcid{0000-0002-8108-9179}\inst{\ref{aff83}}
\and F.~Pasian\orcid{0000-0002-4869-3227}\inst{\ref{aff59}}
\and K.~Pedersen\inst{\ref{aff109}}
\and W.~J.~Percival\orcid{0000-0002-0644-5727}\inst{\ref{aff110},\ref{aff111},\ref{aff112}}
\and V.~Pettorino\inst{\ref{aff107}}
\and S.~Pires\orcid{0000-0002-0249-2104}\inst{\ref{aff45}}
\and G.~Polenta\orcid{0000-0003-4067-9196}\inst{\ref{aff113}}
\and M.~Poncet\inst{\ref{aff64}}
\and L.~A.~Popa\inst{\ref{aff114}}
\and L.~Pozzetti\orcid{0000-0001-7085-0412}\inst{\ref{aff11}}
\and F.~Raison\orcid{0000-0002-7819-6918}\inst{\ref{aff21}}
\and R.~Rebolo\inst{\ref{aff77},\ref{aff115}}
\and A.~Renzi\orcid{0000-0001-9856-1970}\inst{\ref{aff116},\ref{aff84}}
\and J.~Rhodes\orcid{0000-0002-4485-8549}\inst{\ref{aff32}}
\and G.~Riccio\inst{\ref{aff4}}
\and E.~Romelli\orcid{0000-0003-3069-9222}\inst{\ref{aff59}}
\and M.~Roncarelli\orcid{0000-0001-9587-7822}\inst{\ref{aff11}}
\and E.~Rossetti\orcid{0000-0003-0238-4047}\inst{\ref{aff62}}
\and R.~Saglia\orcid{0000-0003-0378-7032}\inst{\ref{aff20},\ref{aff21}}
\and Z.~Sakr\orcid{0000-0002-4823-3757}\inst{\ref{aff117},\ref{aff15},\ref{aff118}}
\and A.~G.~S\'anchez\orcid{0000-0003-1198-831X}\inst{\ref{aff21}}
\and D.~Sapone\orcid{0000-0001-7089-4503}\inst{\ref{aff119}}
\and P.~Schneider\orcid{0000-0001-8561-2679}\inst{\ref{aff103}}
\and T.~Schrabback\orcid{0000-0002-6987-7834}\inst{\ref{aff120}}
\and A.~Secroun\orcid{0000-0003-0505-3710}\inst{\ref{aff86}}
\and G.~Seidel\orcid{0000-0003-2907-353X}\inst{\ref{aff27}}
\and S.~Serrano\orcid{0000-0002-0211-2861}\inst{\ref{aff121},\ref{aff122},\ref{aff123}}
\and C.~Sirignano\orcid{0000-0002-0995-7146}\inst{\ref{aff116},\ref{aff84}}
\and G.~Sirri\orcid{0000-0003-2626-2853}\inst{\ref{aff19}}
\and J.~Skottfelt\orcid{0000-0003-1310-8283}\inst{\ref{aff124}}
\and L.~Stanco\orcid{0000-0002-9706-5104}\inst{\ref{aff84}}
\and J.~Steinwagner\orcid{0000-0001-7443-1047}\inst{\ref{aff21}}
\and P.~Tallada-Cresp\'{i}\orcid{0000-0002-1336-8328}\inst{\ref{aff72},\ref{aff73}}
\and D.~Tavagnacco\orcid{0000-0001-7475-9894}\inst{\ref{aff59}}
\and A.~N.~Taylor\inst{\ref{aff78}}
\and I.~Tereno\inst{\ref{aff81},\ref{aff125}}
\and R.~Toledo-Moreo\orcid{0000-0002-2997-4859}\inst{\ref{aff126}}
\and F.~Torradeflot\orcid{0000-0003-1160-1517}\inst{\ref{aff73},\ref{aff72}}
\and I.~Tutusaus\orcid{0000-0002-3199-0399}\inst{\ref{aff15}}
\and E.~A.~Valentijn\inst{\ref{aff52}}
\and L.~Valenziano\orcid{0000-0002-1170-0104}\inst{\ref{aff11},\ref{aff127}}
\and T.~Vassallo\orcid{0000-0001-6512-6358}\inst{\ref{aff20},\ref{aff59}}
\and Y.~Wang\orcid{0000-0002-4749-2984}\inst{\ref{aff128}}
\and J.~Weller\orcid{0000-0002-8282-2010}\inst{\ref{aff20},\ref{aff21}}
\and E.~Zucca\orcid{0000-0002-5845-8132}\inst{\ref{aff11}}
\and C.~Burigana\orcid{0000-0002-3005-5796}\inst{\ref{aff30},\ref{aff127}}
\and V.~Scottez\inst{\ref{aff105},\ref{aff129}}
\and M.~Viel\orcid{0000-0002-2642-5707}\inst{\ref{aff58},\ref{aff59},\ref{aff61},\ref{aff60},\ref{aff130}}
\and D.~Scott\orcid{0000-0002-6878-9840}\inst{\ref{aff131}}
\and S.~Vegetti\orcid{0009-0006-0592-2882}\inst{\ref{aff2}}}
										   
\institute{Institute of Physics, Laboratory of Astrophysics, Ecole Polytechnique F\'ed\'erale de Lausanne (EPFL), Observatoire de Sauverny, 1290 Versoix, Switzerland\label{aff1}
\and
Max-Planck-Institut f\"ur Astrophysik, Karl-Schwarzschild-Str.~1, 85748 Garching, Germany\label{aff2}
\and
SCITAS, Ecole Polytechnique F\'ed\'erale de Lausanne (EPFL), 1015 Lausanne, Switzerland\label{aff3}
\and
INAF-Osservatorio Astronomico di Capodimonte, Via Moiariello 16, 80131 Napoli, Italy\label{aff4}
\and
Institute of Cosmology and Gravitation, University of Portsmouth, Portsmouth PO1 3FX, UK\label{aff5}
\and
Institut de Ci\`{e}ncies del Cosmos (ICCUB), Universitat de Barcelona (IEEC-UB), Mart\'{i} i Franqu\`{e}s 1, 08028 Barcelona, Spain\label{aff6}
\and
Instituci\'o Catalana de Recerca i Estudis Avan\c{c}ats (ICREA), Passeig de Llu\'{\i}s Companys 23, 08010 Barcelona, Spain\label{aff7}
\and
Aix-Marseille Universit\'e, CNRS, CNES, LAM, Marseille, France\label{aff8}
\and
Institut d'Astrophysique de Paris, UMR 7095, CNRS, and Sorbonne Universit\'e, 98 bis boulevard Arago, 75014 Paris, France\label{aff9}
\and
Dipartimento di Fisica e Astronomia "Augusto Righi" - Alma Mater Studiorum Universit\`a di Bologna, via Piero Gobetti 93/2, 40129 Bologna, Italy\label{aff10}
\and
INAF-Osservatorio di Astrofisica e Scienza dello Spazio di Bologna, Via Piero Gobetti 93/3, 40129 Bologna, Italy\label{aff11}
\and
Department of Physics "E. Pancini", University Federico II, Via Cinthia 6, 80126, Napoli, Italy\label{aff12}
\and
INFN section of Naples, Via Cinthia 6, 80126, Napoli, Italy\label{aff13}
\and
Technical University of Munich, TUM School of Natural Sciences, Physics Department, James-Franck-Str.~1, 85748 Garching, Germany\label{aff14}
\and
Institut de Recherche en Astrophysique et Plan\'etologie (IRAP), Universit\'e de Toulouse, CNRS, UPS, CNES, 14 Av. Edouard Belin, 31400 Toulouse, France\label{aff15}
\and
UCB Lyon 1, CNRS/IN2P3, IUF, IP2I Lyon, 4 rue Enrico Fermi, 69622 Villeurbanne, France\label{aff16}
\and
Department of Astronomy, University of Cape Town, Rondebosch, Cape Town, 7700, South Africa\label{aff17}
\and
STAR Institute, University of Li{\`e}ge, Quartier Agora, All\'ee du six Ao\^ut 19c, 4000 Li\`ege, Belgium\label{aff18}
\and
INFN-Sezione di Bologna, Viale Berti Pichat 6/2, 40127 Bologna, Italy\label{aff19}
\and
Universit\"ats-Sternwarte M\"unchen, Fakult\"at f\"ur Physik, Ludwig-Maximilians-Universit\"at M\"unchen, Scheinerstrasse 1, 81679 M\"unchen, Germany\label{aff20}
\and
Max Planck Institute for Extraterrestrial Physics, Giessenbachstr. 1, 85748 Garching, Germany\label{aff21}
\and
INFN, Sezione di Lecce, Via per Arnesano, CP-193, 73100, Lecce, Italy\label{aff22}
\and
Department of Mathematics and Physics E. De Giorgi, University of Salento, Via per Arnesano, CP-I93, 73100, Lecce, Italy\label{aff23}
\and
INAF-Sezione di Lecce, c/o Dipartimento Matematica e Fisica, Via per Arnesano, 73100, Lecce, Italy\label{aff24}
\and
Department of Physics, Oxford University, Keble Road, Oxford OX1 3RH, UK\label{aff25}
\and
Jodrell Bank Centre for Astrophysics, Department of Physics and Astronomy, University of Manchester, Oxford Road, Manchester M13 9PL, UK\label{aff26}
\and
Max-Planck-Institut f\"ur Astronomie, K\"onigstuhl 17, 69117 Heidelberg, Germany\label{aff27}
\and
Department of Physics, Centre for Extragalactic Astronomy, Durham University, South Road, Durham, DH1 3LE, UK\label{aff28}
\and
Department of Physics, Institute for Computational Cosmology, Durham University, South Road, Durham, DH1 3LE, UK\label{aff29}
\and
INAF, Istituto di Radioastronomia, Via Piero Gobetti 101, 40129 Bologna, Italy\label{aff30}
\and
University of Applied Sciences and Arts of Northwestern Switzerland, School of Engineering, 5210 Windisch, Switzerland\label{aff31}
\and
Jet Propulsion Laboratory, California Institute of Technology, 4800 Oak Grove Drive, Pasadena, CA, 91109, USA\label{aff32}
\and
Centro de Astrof\'{\i}sica da Universidade do Porto, Rua das Estrelas, 4150-762 Porto, Portugal\label{aff33}
\and
Instituto de Astrof\'isica e Ci\^encias do Espa\c{c}o, Universidade do Porto, CAUP, Rua das Estrelas, PT4150-762 Porto, Portugal\label{aff34}
\and
School of Physical Sciences, The Open University, Milton Keynes, MK7 6AA, UK\label{aff35}
\and
Minnesota Institute for Astrophysics, University of Minnesota, 116 Church St SE, Minneapolis, MN 55455, USA\label{aff36}
\and
Dipartimento di Fisica "Aldo Pontremoli", Universit\`a degli Studi di Milano, Via Celoria 16, 20133 Milano, Italy\label{aff37}
\and
INAF-IASF Milano, Via Alfonso Corti 12, 20133 Milano, Italy\label{aff38}
\and
Department of Physics and Astronomy, University of the Western Cape, Bellville, Cape Town, 7535, South Africa\label{aff39}
\and
Observatoire de Sauverny, Ecole Polytechnique F\'ed\'erale de Lausanne, 1290 Versoix, Switzerland\label{aff40}
\and
David A. Dunlap Department of Astronomy \& Astrophysics, University of Toronto, 50 St George Street, Toronto, Ontario M5S 3H4, Canada\label{aff41}
\and
Laboratoire d'Astrophysique de Bordeaux, CNRS and Universit\'e de Bordeaux, All\'ee Geoffroy St. Hilaire, 33165 Pessac, France\label{aff42}
\and
Institut universitaire de France (IUF), 1 rue Descartes, 75231 PARIS CEDEX 05, France\label{aff43}
\and
Caltech/IPAC, 1200 E. California Blvd., Pasadena, CA 91125, USA\label{aff44}
\and
Universit\'e Paris-Saclay, Universit\'e Paris Cit\'e, CEA, CNRS, AIM, 91191, Gif-sur-Yvette, France\label{aff45}
\and
Universit\'e de Strasbourg, CNRS, Observatoire astronomique de Strasbourg, UMR 7550, 67000 Strasbourg, France\label{aff46}
\and
Department of Physics, Universit\'{e} de Montr\'{e}al, 2900 Edouard Montpetit Blvd, Montr\'{e}al, Qu\'{e}bec H3T 1J4, Canada\label{aff47}
\and
Instituto de F\'isica de Cantabria, Edificio Juan Jord\'a, Avenida de los Castros, 39005 Santander, Spain\label{aff48}
\and
Dipartimento di Fisica e Scienze della Terra, Universit\`a degli Studi di Ferrara, Via Giuseppe Saragat 1, 44122 Ferrara, Italy\label{aff49}
\and
Departamento F\'isica Aplicada, Universidad Polit\'ecnica de Cartagena, Campus Muralla del Mar, 30202 Cartagena, Murcia, Spain\label{aff50}
\and
Laboratoire univers et particules de Montpellier, Universit\'e de Montpellier, CNRS, 34090 Montpellier, France\label{aff51}
\and
Kapteyn Astronomical Institute, University of Groningen, PO Box 800, 9700 AV Groningen, The Netherlands\label{aff52}
\and
ESAC/ESA, Camino Bajo del Castillo, s/n., Urb. Villafranca del Castillo, 28692 Villanueva de la Ca\~nada, Madrid, Spain\label{aff53}
\and
INAF-Osservatorio Astrofisico di Arcetri, Largo E. Fermi 5, 50125, Firenze, Italy\label{aff54}
\and
Institute of Physics, Laboratory for Galaxy Evolution, Ecole Polytechnique F\'ed\'erale de Lausanne, Observatoire de Sauverny, CH-1290 Versoix, Switzerland\label{aff55}
\and
School of Mathematics and Physics, University of Surrey, Guildford, Surrey, GU2 7XH, UK\label{aff56}
\and
INAF-Osservatorio Astronomico di Brera, Via Brera 28, 20122 Milano, Italy\label{aff57}
\and
IFPU, Institute for Fundamental Physics of the Universe, via Beirut 2, 34151 Trieste, Italy\label{aff58}
\and
INAF-Osservatorio Astronomico di Trieste, Via G. B. Tiepolo 11, 34143 Trieste, Italy\label{aff59}
\and
INFN, Sezione di Trieste, Via Valerio 2, 34127 Trieste TS, Italy\label{aff60}
\and
SISSA, International School for Advanced Studies, Via Bonomea 265, 34136 Trieste TS, Italy\label{aff61}
\and
Dipartimento di Fisica e Astronomia, Universit\`a di Bologna, Via Gobetti 93/2, 40129 Bologna, Italy\label{aff62}
\and
INAF-Osservatorio Astronomico di Padova, Via dell'Osservatorio 5, 35122 Padova, Italy\label{aff63}
\and
Centre National d'Etudes Spatiales -- Centre spatial de Toulouse, 18 avenue Edouard Belin, 31401 Toulouse Cedex 9, France\label{aff64}
\and
INAF-Osservatorio Astrofisico di Torino, Via Osservatorio 20, 10025 Pino Torinese (TO), Italy\label{aff65}
\and
Dipartimento di Fisica, Universit\`a di Genova, Via Dodecaneso 33, 16146, Genova, Italy\label{aff66}
\and
INFN-Sezione di Genova, Via Dodecaneso 33, 16146, Genova, Italy\label{aff67}
\and
Faculdade de Ci\^encias da Universidade do Porto, Rua do Campo de Alegre, 4150-007 Porto, Portugal\label{aff68}
\and
Dipartimento di Fisica, Universit\`a degli Studi di Torino, Via P. Giuria 1, 10125 Torino, Italy\label{aff69}
\and
INFN-Sezione di Torino, Via P. Giuria 1, 10125 Torino, Italy\label{aff70}
\and
Mullard Space Science Laboratory, University College London, Holmbury St Mary, Dorking, Surrey RH5 6NT, UK\label{aff71}
\and
Centro de Investigaciones Energ\'eticas, Medioambientales y Tecnol\'ogicas (CIEMAT), Avenida Complutense 40, 28040 Madrid, Spain\label{aff72}
\and
Port d'Informaci\'{o} Cient\'{i}fica, Campus UAB, C. Albareda s/n, 08193 Bellaterra (Barcelona), Spain\label{aff73}
\and
Institute for Theoretical Particle Physics and Cosmology (TTK), RWTH Aachen University, 52056 Aachen, Germany\label{aff74}
\and
INAF-Osservatorio Astronomico di Roma, Via Frascati 33, 00078 Monteporzio Catone, Italy\label{aff75}
\and
Dipartimento di Fisica e Astronomia "Augusto Righi" - Alma Mater Studiorum Universit\`a di Bologna, Viale Berti Pichat 6/2, 40127 Bologna, Italy\label{aff76}
\and
Instituto de Astrof\'{\i}sica de Canarias, V\'{\i}a L\'actea, 38205 La Laguna, Tenerife, Spain\label{aff77}
\and
Institute for Astronomy, University of Edinburgh, Royal Observatory, Blackford Hill, Edinburgh EH9 3HJ, UK\label{aff78}
\and
European Space Agency/ESRIN, Largo Galileo Galilei 1, 00044 Frascati, Roma, Italy\label{aff79}
\and
Universit\'e Claude Bernard Lyon 1, CNRS/IN2P3, IP2I Lyon, UMR 5822, Villeurbanne, F-69100, France\label{aff80}
\and
Departamento de F\'isica, Faculdade de Ci\^encias, Universidade de Lisboa, Edif\'icio C8, Campo Grande, PT1749-016 Lisboa, Portugal\label{aff81}
\and
Instituto de Astrof\'isica e Ci\^encias do Espa\c{c}o, Faculdade de Ci\^encias, Universidade de Lisboa, Campo Grande, 1749-016 Lisboa, Portugal\label{aff82}
\and
Department of Astronomy, University of Geneva, ch. d'Ecogia 16, 1290 Versoix, Switzerland\label{aff83}
\and
INFN-Padova, Via Marzolo 8, 35131 Padova, Italy\label{aff84}
\and
INAF-Istituto di Astrofisica e Planetologia Spaziali, via del Fosso del Cavaliere, 100, 00100 Roma, Italy\label{aff85}
\and
Aix-Marseille Universit\'e, CNRS/IN2P3, CPPM, Marseille, France\label{aff86}
\and
Istituto Nazionale di Fisica Nucleare, Sezione di Bologna, Via Irnerio 46, 40126 Bologna, Italy\label{aff87}
\and
FRACTAL S.L.N.E., calle Tulip\'an 2, Portal 13 1A, 28231, Las Rozas de Madrid, Spain\label{aff88}
\and
Institute of Theoretical Astrophysics, University of Oslo, P.O. Box 1029 Blindern, 0315 Oslo, Norway\label{aff89}
\and
Leiden Observatory, Leiden University, Einsteinweg 55, 2333 CC Leiden, The Netherlands\label{aff90}
\and
Department of Physics, Lancaster University, Lancaster, LA1 4YB, UK\label{aff91}
\and
Felix Hormuth Engineering, Goethestr. 17, 69181 Leimen, Germany\label{aff92}
\and
Technical University of Denmark, Elektrovej 327, 2800 Kgs. Lyngby, Denmark\label{aff93}
\and
Cosmic Dawn Center (DAWN), Denmark\label{aff94}
\and
NASA Goddard Space Flight Center, Greenbelt, MD 20771, USA\label{aff95}
\and
Department of Physics and Astronomy, University College London, Gower Street, London WC1E 6BT, UK\label{aff96}
\and
Department of Physics and Helsinki Institute of Physics, Gustaf H\"allstr\"omin katu 2, 00014 University of Helsinki, Finland\label{aff97}
\and
Universit\'e de Gen\`eve, D\'epartement de Physique Th\'eorique and Centre for Astroparticle Physics, 24 quai Ernest-Ansermet, CH-1211 Gen\`eve 4, Switzerland\label{aff98}
\and
Department of Physics, P.O. Box 64, 00014 University of Helsinki, Finland\label{aff99}
\and
Helsinki Institute of Physics, Gustaf H{\"a}llstr{\"o}min katu 2, University of Helsinki, Helsinki, Finland\label{aff100}
\and
NOVA optical infrared instrumentation group at ASTRON, Oude Hoogeveensedijk 4, 7991PD, Dwingeloo, The Netherlands\label{aff101}
\and
Centre de Calcul de l'IN2P3/CNRS, 21 avenue Pierre de Coubertin 69627 Villeurbanne Cedex, France\label{aff102}
\and
Universit\"at Bonn, Argelander-Institut f\"ur Astronomie, Auf dem H\"ugel 71, 53121 Bonn, Germany\label{aff103}
\and
INFN-Sezione di Roma, Piazzale Aldo Moro, 2 - c/o Dipartimento di Fisica, Edificio G. Marconi, 00185 Roma, Italy\label{aff104}
\and
Institut d'Astrophysique de Paris, 98bis Boulevard Arago, 75014, Paris, France\label{aff105}
\and
Institut de F\'{i}sica d'Altes Energies (IFAE), The Barcelona Institute of Science and Technology, Campus UAB, 08193 Bellaterra (Barcelona), Spain\label{aff106}
\and
European Space Agency/ESTEC, Keplerlaan 1, 2201 AZ Noordwijk, The Netherlands\label{aff107}
\and
School of Mathematics, Statistics and Physics, Newcastle University, Herschel Building, Newcastle-upon-Tyne, NE1 7RU, UK\label{aff108}
\and
DARK, Niels Bohr Institute, University of Copenhagen, Jagtvej 155, 2200 Copenhagen, Denmark\label{aff109}
\and
Waterloo Centre for Astrophysics, University of Waterloo, Waterloo, Ontario N2L 3G1, Canada\label{aff110}
\and
Department of Physics and Astronomy, University of Waterloo, Waterloo, Ontario N2L 3G1, Canada\label{aff111}
\and
Perimeter Institute for Theoretical Physics, Waterloo, Ontario N2L 2Y5, Canada\label{aff112}
\and
Space Science Data Center, Italian Space Agency, via del Politecnico snc, 00133 Roma, Italy\label{aff113}
\and
Institute of Space Science, Str. Atomistilor, nr. 409 M\u{a}gurele, Ilfov, 077125, Romania\label{aff114}
\and
Universidad de La Laguna, Departamento de Astrof\'{\i}sica, 38206 La Laguna, Tenerife, Spain\label{aff115}
\and
Dipartimento di Fisica e Astronomia "G. Galilei", Universit\`a di Padova, Via Marzolo 8, 35131 Padova, Italy\label{aff116}
\and
Institut f\"ur Theoretische Physik, University of Heidelberg, Philosophenweg 16, 69120 Heidelberg, Germany\label{aff117}
\and
Universit\'e St Joseph; Faculty of Sciences, Beirut, Lebanon\label{aff118}
\and
Departamento de F\'isica, FCFM, Universidad de Chile, Blanco Encalada 2008, Santiago, Chile\label{aff119}
\and
Universit\"at Innsbruck, Institut f\"ur Astro- und Teilchenphysik, Technikerstr. 25/8, 6020 Innsbruck, Austria\label{aff120}
\and
Institut d'Estudis Espacials de Catalunya (IEEC),  Edifici RDIT, Campus UPC, 08860 Castelldefels, Barcelona, Spain\label{aff121}
\and
Satlantis, University Science Park, Sede Bld 48940, Leioa-Bilbao, Spain\label{aff122}
\and
Institute of Space Sciences (ICE, CSIC), Campus UAB, Carrer de Can Magrans, s/n, 08193 Barcelona, Spain\label{aff123}
\and
Centre for Electronic Imaging, Open University, Walton Hall, Milton Keynes, MK7~6AA, UK\label{aff124}
\and
Instituto de Astrof\'isica e Ci\^encias do Espa\c{c}o, Faculdade de Ci\^encias, Universidade de Lisboa, Tapada da Ajuda, 1349-018 Lisboa, Portugal\label{aff125}
\and
Universidad Polit\'ecnica de Cartagena, Departamento de Electr\'onica y Tecnolog\'ia de Computadoras,  Plaza del Hospital 1, 30202 Cartagena, Spain\label{aff126}
\and
INFN-Bologna, Via Irnerio 46, 40126 Bologna, Italy\label{aff127}
\and
Infrared Processing and Analysis Center, California Institute of Technology, Pasadena, CA 91125, USA\label{aff128}
\and
ICL, Junia, Universit\'e Catholique de Lille, LITL, 59000 Lille, France\label{aff129}
\and
ICSC - Centro Nazionale di Ricerca in High Performance Computing, Big Data e Quantum Computing, Via Magnanelli 2, Bologna, Italy\label{aff130}
\and
Department of Physics and Astronomy, University of British Columbia, Vancouver, BC V6T 1Z1, Canada\label{aff131}}    


 \abstract{
 We investigated the ability of the \Euclid telescope to detect galaxy-scale gravitational lenses. To do so, we performed a systematic visual inspection of the 0.7\,deg$^2$ \Euclid Early Release Observations data towards the Perseus cluster using both the high-resolution \IE band and the lower-resolution \YE, \JE, and \HE bands.
 Each extended source brighter than magnitude~23 in \IE  was inspected by 41 expert human classifiers.
 This amounts to 12\,086 stamps of \ang{;;10}\,$\times$\,\ang{;;10}.
 We found 3 grade A and 13 grade B candidates.
 We assessed the validity of these 16 candidates by modelling them and checking that they are consistent with a single source lensed by a plausible mass distribution. Five of the candidates pass this check, five others are rejected by the modelling, and six are inconclusive.
Extrapolating from the five successfully modelled candidates, we infer that the full $14\,000\,{\rm deg}^2$ of the \Euclid Wide Survey should contain $100\,000^{+70\,000}_{-30\,000}$ galaxy-galaxy lenses that are both discoverable through visual inspection and have valid lens models.
This is consistent with theoretical forecasts of \num{170000} discoverable galaxy-galaxy lenses in \Euclid.
Our five modelled lenses have Einstein radii in the range $\ang{;;0.68}\,<\,\theta_\mathrm{E}\,<\ang{;;1.24}$, but their Einstein radius distribution is on the higher side when compared to theoretical forecasts.
This suggests that our methodology is likely missing small-Einstein-radius systems.
Whilst it is implausible to visually inspect the full \Euclid dataset, our results corroborate the promise that \Euclid will ultimately deliver a sample of around $10^5$ galaxy-scale lenses.
 }

      \title{\vspace{-0.09cm}\Euclid: The Early Release Observations Lens Search Experiment\thanks{This paper is published on behalf of the Euclid Consortium.}}
      
    \maketitle

\section{Introduction}

Strong gravitational lensing by massive galaxies offers a plethora of applications in both  cosmology and astrophysics.
Some notable examples include measuring the total mass of lens galaxies within the Einstein radius and disentangling the contributions from visible and dark matter components \citep[e.g.][]{Auger2009}.
When coupled with deep spectroscopic observations, it enables the placement of constraints on the stellar initial mass function of lens galaxies \citep[e.g.][]{Ferreras2010, Dutton2014, Sonnenfeld2019}.
Thanks to the lensing magnification, strong gravitational lensing serves as a natural telescope to study lensed sources otherwise too faint or angularly too small to be detected \citep[e.g.][]{Hezaveh2013}, and even to map their velocity field with unprecedented spatial resolution \citep[e.g.][]{Paraficz2018}.
Studying small-scale distortions of lensed images or arcs allows us to infer the presence of low-mass dark halos either in the lens or along the line of sight up to the source redshift, and in turn to study the properties of dark matter \citep[e.g.][]{Vegetti2010, Riordan2023, 2024MNRAS.tmp.1779G}.
Moreover, when the lensed source is time-variable, such as a quasar or a supernova, strong lensing offers an independent way of measuring the expansion rate of the Universe, $H_0$ \citep{Refsdal1964, Wong2020, Shajib2023, 2024A&A...684L..23G, 2025ApJ...979...13P}. 

Lastly, the measurement of weak lensing shear from strong lensing images has been proposed as a potential cosmological probe \citep{2017JCAP...04..049B, 2018ApJ...852L..14B}. A minimal model for the shear that is non-degenerate with lens model parameters was derived by \cite{2021JCAP...08..024F}, and this quantity was shown to be measurable in mock imaging data by \cite{2023MNRAS.520.5982H}. For high-precision cosmological constraints, however, a \Euclid-sized dataset of $\mathcal{O}(10^5)$ strong lenses will be required.

Galaxy-scale strong lensing events are rare, with about one object out of thousands showing lensing features \citep{OM10,Collett2015}. In addition, these systems are small in angular terms, spanning between tenths of an arcsecond to a few arcseconds on the sky, and the lensed images of the sources (arcs, rings, and multiple point sources) are often hidden in the glare of the foreground lensing galaxy. Because lenses are rare, compact, and low-contrast objects, they are best discovered in deep, sharp, wide-field surveys. \Euclid is exactly that \citep{EuclidSkyOverview}, and is the focus of the present work.
In fact, lensed quasi-stellar object candidates have already been proposed from \Euclid observations \citep{EROPerseusOverview}.

However, even with high-quality \Euclid data, finding lenses remains challenging, not only because of the intrinsic rarity of strong lensing, but also because other non-lensing objects mimic the morphology of lenses (e.g. ring galaxies, galaxy mergers, spiral galaxies, or even random alignments).
So far, the best methods for discovering lenses involve various flavours of convolutional neural networks (see \citealt{Petrillo2019, Jacobs2019, Stein2022, Savary2022, Rojas2022}, to cite just a few) and, more recently, transformer networks \citep{Thuruthipilly22, Grespan_Thuruthipilly24, 2025arXiv250115679G}.
These networks require training sets that match the instrumental and astrophysical properties of the data very closely.
Since we do not possess enough strong lens observations to constitute a training set, they have to be simulated, and the effectiveness of simulations in emulating nature is limited.

Machine learning methods are effective at finding lenses, but they are limited in terms of purity and false-positive rates. When false-positive rates are below 1\%, such methods are useful for pre-selecting candidates from parent samples of tens or even hundreds of millions of galaxies \citep{Canameras2023}.
However, as the true -- and poorly known -- prevalence of lenses on the plane of the sky is very low, even a 1\% false-positive rate can result in samples dominated by contaminants and unconvincing candidates.
This is known as the base rate fallacy.
As a result, human visual inspection is almost always necessary to fine-tune the network selection, although humans themselves sometimes have difficulty deciding on the validity of a lens candidate.
This is especially true for small-Einstein-radius lenses.

Human visual inspection is useful on its own as a lens finding method but is limited by the volume of data that can be inspected by a team of experts in a reasonable amount of time.
However, it has the advantage that it can pick up unusual lensing configurations \citep[e.g.][]{Keeton2009, OrbandeXivry2009, CollettBacon16} that, by definition, do not appear in large numbers in simulated training sets.
It is also the only way to evaluate the prevalence of lenses on the plane of the sky and therefore the expected surface density of lensing systems for a given instrument, depth, spatial resolution, and wavelength. This has been attempted a few times in the past, for example in \textit{Hubble} Space Telescope imaging \citep[e.g.][]{Faure2008, Pawase2014,2022A&A...667A.141G}.
\citet{Rojas2023} also evaluated the performance of visual inspection with simulated data. 

In the \ac{ERO} Lens Search Experiment (ELSE), we carried out a blind visual search of galaxy-scale strong lensing systems using some of the first data from the European Space Agency (ESA) satellite \Euclid \citep{Scaramella-EP1,EuclidSkyOverview}.
We focused on the \Euclid \ac{ERO} imaging of the Perseus galaxy cluster (\citealt{EROData}), and the only selection criterion for the extended sources to be inspected was a magnitude cut.
Our search is therefore one of the broadest visual searches carried out so far in terms of pre-selection.
The goals were:
(1) to evaluate the performance of the \Euclid telescope at finding galaxy-scale lenses;
(2) to study the prevalence of strong lenses found by humans in \Euclid;
(3) to test the efficiency of human experts at finding lenses;
and (4) to optimise future visual inspections that will be performed on lens candidates found by the automatic pipeline in \Euclid.

The humans involved in this exercise are all lensing experts but were limited in number (41 exactly).
Our work therefore contrasts with more intensive citizen science searches involving much larger numbers of humans but spanning a much smaller range in terms of expertise \citep{Marshall2015, More2016}. 

Finally, assuming subtraction of the lens light, \Euclid is predicted to find \num{170000} lenses in its six-year main survey \citep{Collett2015}.
We tested that prediction in this work, with the caveat that we did not perform any subtraction of the lens light.
Thus, our results represent a lower bound in this regard.

We introduce the \Euclid \ac{ERO} observations of the Perseus cluster in \cref{sec:data}.
We then explain the methodology of the visual inspections in \cref{sec:method}, along with the visualisation tools used.
Afterwards, we report the results from the visual inspection in \cref{sec:visual_inspection_results} and present the sample of lens candidates and some initial modelling in \cref{sec:results}.
In \cref{sec:discussion} we then compare the sample against the literature and estimate the number of lenses that will be discovered in the \ac{EWS}. We conclude in \cref{sec:conclusions}.

\section{Data}
\label{sec:data}
\begin{figure*}[t!]
\centering
\includegraphics[width=0.99\textwidth]{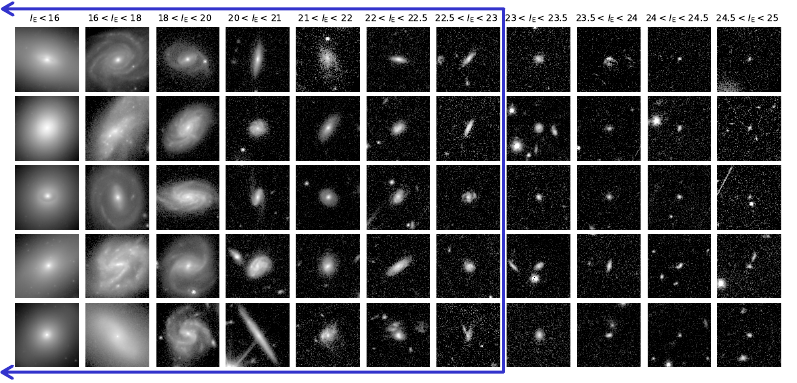}
\caption{$\ang{;;9.9}\times\ang{;;9.9}$  VIS cutouts of galaxies randomly selected across different $\IE$ bins, ranging from the brightest galaxies with $\IE < 16$ (leftmost) down to very faint systems with $24.5 < \IE < 25$ (rightmost).
The blue arrow represents the magnitude cut used to select our parent sample.}
\label{fig:images_vs_magnitude}
\end{figure*}

The \Euclid \ac{ERO} programme \citep{EROcite} targeted the Perseus galaxy cluster, obtaining very deep data of the central region of the cluster in $0.7\,{\rm deg^2}$, in the broad optical filter \IE from the \ac{VIS} instrument \citep{EuclidSkyVIS}, and the three broad filters \YE, \JE, and \HE in the near-infrared from the \ac{NISP} instrument \citep{EuclidSkyNISP}.

These data were collected during the \Euclid performance verification phase in September 2023 \citep{EROData}.
All \Euclid science observations adhere to a reference observing sequences \citep[ROSs;][]{Scaramella-EP1}, which consists of four dithered exposures lasting $566$ seconds each in the \IE filter, and four dithered exposures of $87.2$ seconds each in the \YE, \JE, and \HE filters. 
Four ROSs were obtained for this field, with a total integration time of
\num{7456.0} seconds in the \IE filter and \num{1392.2} seconds in the \YE, \JE, and \HE filters, achieving a depth $0.75$ magnitudes deeper than that of the \ac{EWS}, which relies on a single ROS \citep{Scaramella-EP1}. Therefore, these exceptional data reach a point-source depth of $\IE = 27.3$ ($\YE,\,\JE,\,\HE=24.9$) at $10\,\sigma$, with a \mbox{\ang{;;0.16}} (\mbox{\ang{;;0.48}})  full width at half maximum, and a surface brightness limit of $30.1\,(29.2)\,\mathrm{mag}\,\mathrm{arcsec}^{-2}$.
We refer the reader to \cite{EROData} for more details on the data reduction.

The astrometrically and photometrically calibrated imaging stacks across all four \Euclid bands are accompanied by catalogues for compact sources produced using the tool \texttt{SourceExtractor} \citep{Bertin_Arnouts96} on the `flattened' stack, that is, with the low spatial frequencies removed (sky background and any Galactic cirrus or nebulae).
Multi-wavelength catalogues including VIS, NISP, and ground-based Canada-France-Hawaii Telescope MegaCam photometry have been provided in \cite{EROPerseusOverview}.
After star--galaxy separation and limiting the analysis to sources with $\IE < 23$, we ended up with a parent sample of \mbox{\num{12086}} objects, which was later used in the visual classification procedure.
As shown in \cref{fig:images_vs_magnitude}, the adopted magnitude cut is justified by the fact that sources at fainter magnitudes are compact and featureless.
For the experiment, cutouts in the four \Euclid bands of $\ang{;;9.9} \times \ang{;;9.9}$ (i.e. $99 \times 99$ pixels and $33 \times 33$ pixels in the VIS and NISP bands, respectively) were created using the flattened stacks.

\section{Method}
\label{sec:method}
\subsection{The visualisation tools}
\label{sec:visualization_tools}

\begin{figure*}[t]
\centering
\includegraphics[width=0.61\textwidth]{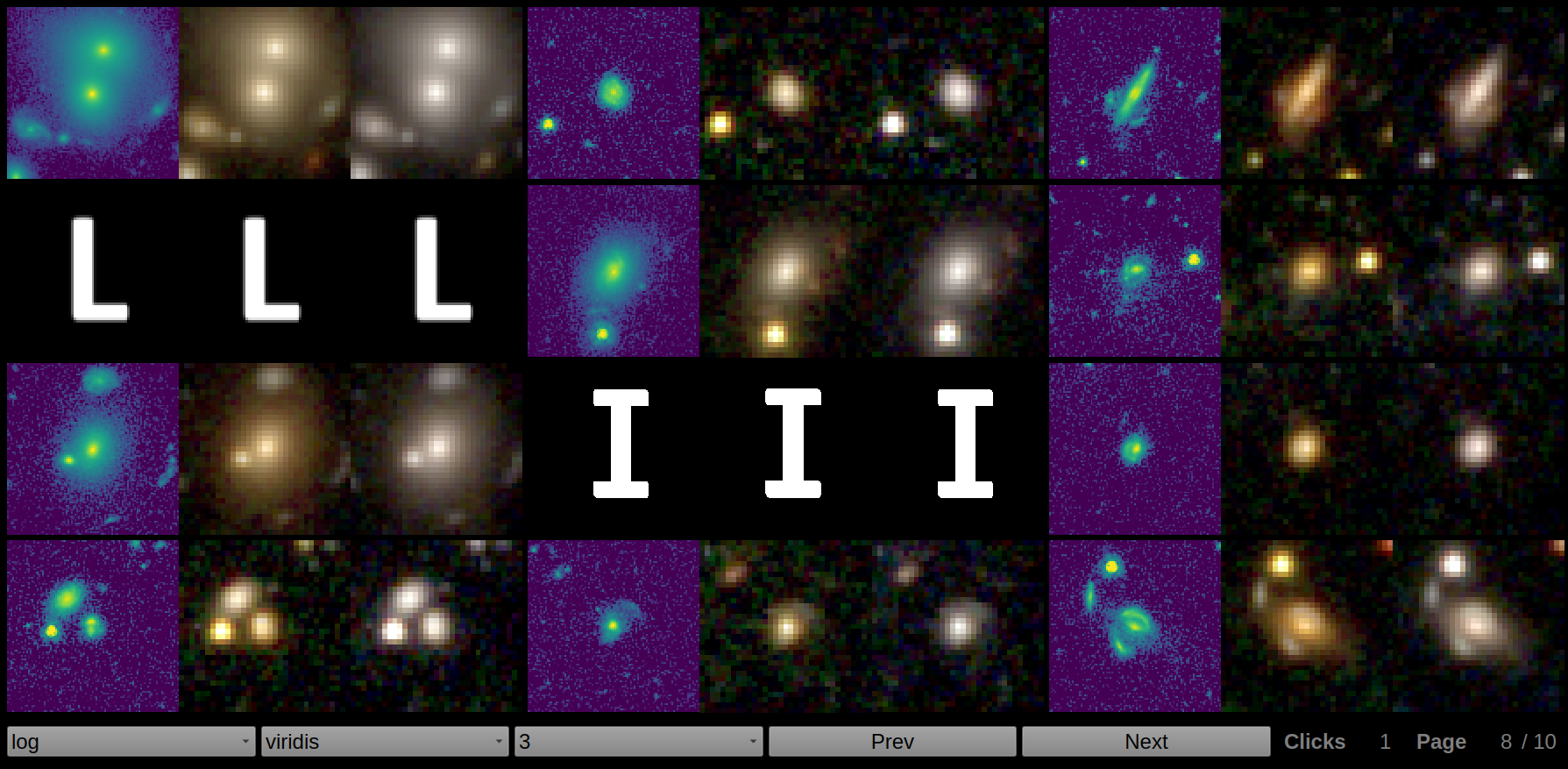}
\includegraphics[width=0.373\textwidth]{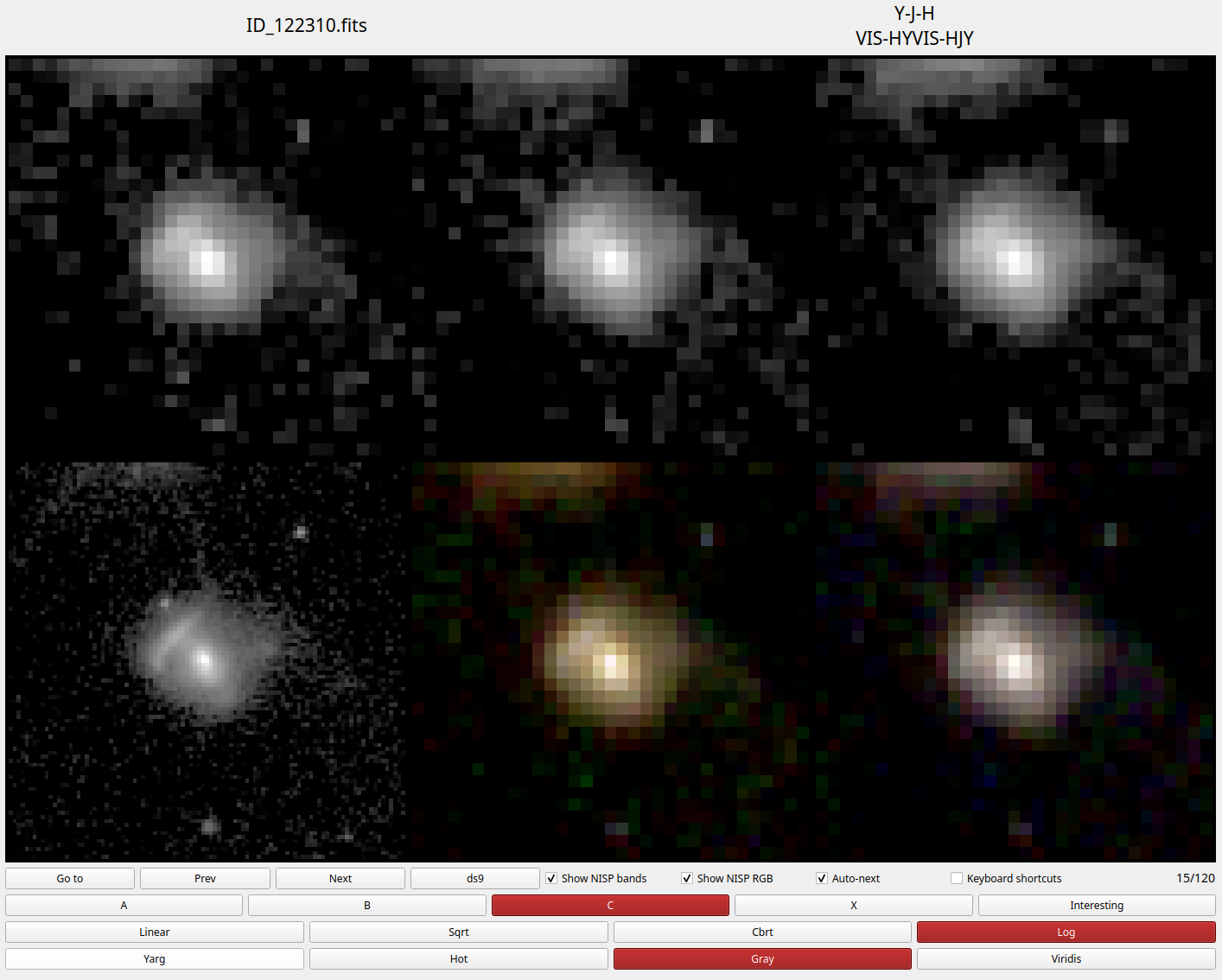}
\caption{Visualisation tools used for the visual inspection.
\emph{Left panel}: Mosaic tool showing 12 sources in a 3~$\times$~4 rectangular grid. The first object of the second row is marked as a lens candidate, whereas the second object of the third row is marked as interesting.
\emph{Right panel}: One-by-one sequential tool showing an object graded as C.
Both tools show a monochromatic high-resolution \IE band image, an \HE\YE\IE RGB composite image, and a \HE\JE\YE RGB composite image (in the second row for the one-by-one sequential tool).
Both RGB composite images are at the NISP resolution.
The one-by-one sequential tool also shows, in its first row, the three NISP bands: \YE, \JE, and \HE.
Users are only required to inspect the high-resolution \IE monochromatic image and the \HE\YE\IE composite image.
}
\label{fig:visualization_tool_extended}
\end{figure*}

We used the visualisation tools prepared by \citet{2025arXiv250310610A} to carry out the visual inspection.\footnote{The version used for this work is available at \url{https://github.com/ClarkGuilty/Qt-stamp-visualizer/tree/ERO_edition}}
The tools are based on the visualisation tools used in \cite{Savary2022} and \cite{Rojas2022}, but re-implemented in the \texttt{Qt} 6 framework with extended functionality targeting the requirements of the \Euclid \ac{ERO} data.
The two applications correspond to a mosaic viewer and a one-by-one sequential viewer, both presented in \cref{fig:visualization_tool_extended}.
The mosaic viewer displays rectangular mosaics of objects for the user to inspect. 
Then, the user is tasked with clicking on objects that show any signs of lensing.
Additionally, the user is allowed to mark objects as `interesting, but not a lens'.
This is exemplified in the left panel of \cref{fig:visualization_tool_extended}.
By contrast, the one-by-one sequential tool displays one object at the time and the user is tasked with classifying it into one of the following non-overlapping categories: 
\bi
\item `A' indicates a sure lens: it shows clear lensing features and no additional information is needed.
\item `B' indicates a probable lens: it shows lensing features but additional information is required to verify it as a definite lens.\item `C' indicates a possible lens: it shows lensing features, but they can be explained without resorting to gravitational lensing.
\item `X' indicates it is definitively not a lens.
\item `Interesting' indicates it is definitively not a lens but is interesting in some other way.
\ei
The grades A, B, and C are deemed as positive grades, whereas X and Interesting are deemed as negative.
The one-by-one sequential tool is shown in the right panel of \cref{fig:visualization_tool_extended}.
Both tools allow the users to change the colour map used for the monochromatic images, as well as the function used to scale the pixel intensities.
For every object to inspect, the tools generate three high-contrast images:
\begin{itemize}
    \item a monochromatic \IE image at the VIS resolution of $\ang{;;0.1}\,{\rm pixel}^{-1}$;
    \item a red-green-blue (RGB) composite image using the \HE, \YE, and \IE bands, at the NISP resolution of $\ang{;;0.3}\,{\rm pixel}^{-1}$;
    \item an RGB composite image using the \HE, \JE, and \YE bands, at the NISP resolution of $\ang{;;0.3}\,{\rm pixel}^{-1}$.
\end{itemize}
Before creating the composite images, we re-projected the \IE band data from the sky-coordinate system and resolution of the VIS instrument to the corresponding ones in the NISP instrument.
This aligns the images and corrects for the different pixel scales between instruments.
The experts were required to use the monochromatic high-resolution \IE band, and the \HE\YE\IE composite.
The \HE\JE\YE composite is not shown by default in the tools, but it is also available.
Additionally, the users also have access to the individual NISP bands when using the one-by-one sequential tool.
This is shown in \cref{fig:visualization_tool_extended}.

\subsection{The visual inspections}
\label{sec:visual_inspections}
\begin{figure*}[t]
\centering
\includegraphics[width=1.0\textwidth]{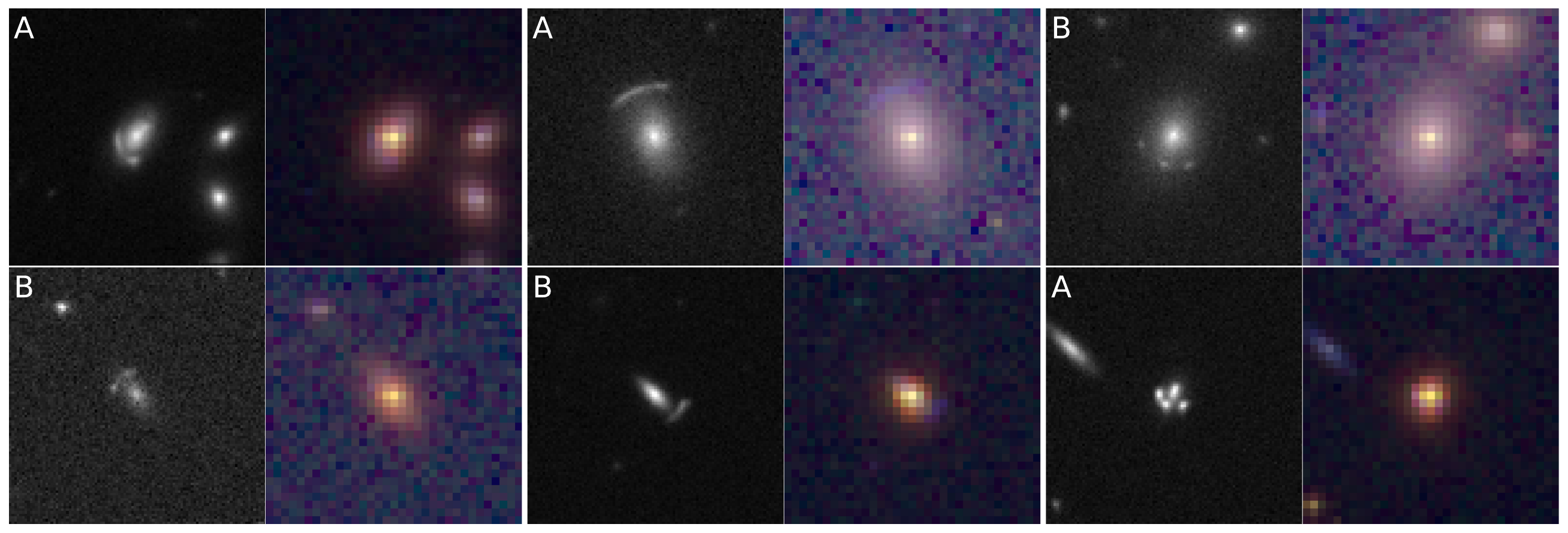}
\caption{Six simulated lenses as seen in the visualisation tools.
The left side of each panel corresponds to the high-resolution \IE band data, and the right to the \HE\YE\IE composite image.
The letter at the upper-left corner of each panel corresponds to the final joint grade given in the visual inspection.
}
\label{fig:all_mocks}
\end{figure*}

To make better use of the large number of strong lensing experts available for the visual inspection, we split the 41 experts into two non-overlapping teams.
This allowed us to try different visual inspection schemes while still retaining enough experts per team for the classifications not to be dominated by noise.
Team~1 comprises 22 experts, and Team~2 comprises 19.
Both teams inspect the \mbox{\num{12086}} cutouts described in \cref{sec:data}, along with six extra cutouts of simulated lenses produced following the prescription of Euclid Consortium, Metcalf et al. (in prep).
The introduction of simulated mocks tests the ability of the experts to identify lenses, and serves as a sanity check of their performance.
The experts are not informed about the mocks in order to avoid any biases.
The six simulated lenses can be seen in \cref{fig:all_mocks}.
Both teams are given two weeks per stage.
There is no communication either between or within the teams, and their results are not combined until after the visual inspections are completed.

Both teams followed a two-stage approach.
First, they focused on cleaning the parent sample of obvious non-lens contaminants.
Then, they inspected  the remaining sources in detail and settled on a final classification following the grades introduced in \cref{sec:visualization_tools}.

Team~1 first inspects the entire parent sample using the mosaic tool.
This corresponds to \mbox{\num{12092}} cutouts after adding the mock lenses.
The experts were allowed to change the number of sources per page shown in the mosaic, with most experts observing between 20 and 42 sources per mosaic page. 
Afterwards, Team~1 reinspected all the objects selected in the first stage  using the one-by-one sequential tool and assigning detailed classifications.

Simultaneously, Team~2 is further divided into three groups: two groups comprising six experts each and one group comprising seven.
Each group inspects one third of the sources using the one-by-one sequential tool, while focusing on rejecting obvious contaminants.
During the first stage, no detailed classification is required.
Subsequently, all the selected sources are reinspected by Team~2 as a whole, using again the one-by-one sequential tool but assigning detailed classifications.

Afterwards, we aggregated the final classifications of each team independently.
The output is a single final classification per team for every source.
Thus, if more than half of the experts voted negatively, then the source was classified as a non-lens;
otherwise, we classified it as the majority vote between A, B, and C.
In the case of a tie between A and B, or A and C, A took precedence.
This prioritisation reflects the expectation that experts will only vote for A if they are confident about the presence of lensing features.
If there is a tie between B and C, we favoured C to minimise noise in the final B sample.
In the event of a tie among all three positive classes, B was preferred.

\section{Visual inspection}
\label{sec:visual_inspection_results}
The experts are anonymised at every stage of the analysis to avoid biasing the visual inspection results.
We refer to the experts only by their expert ID, for example, `expert number 12'. 

\subsection{Team 1} 
\begin{figure}[t!]
\centering
\includegraphics[width=0.5\textwidth]{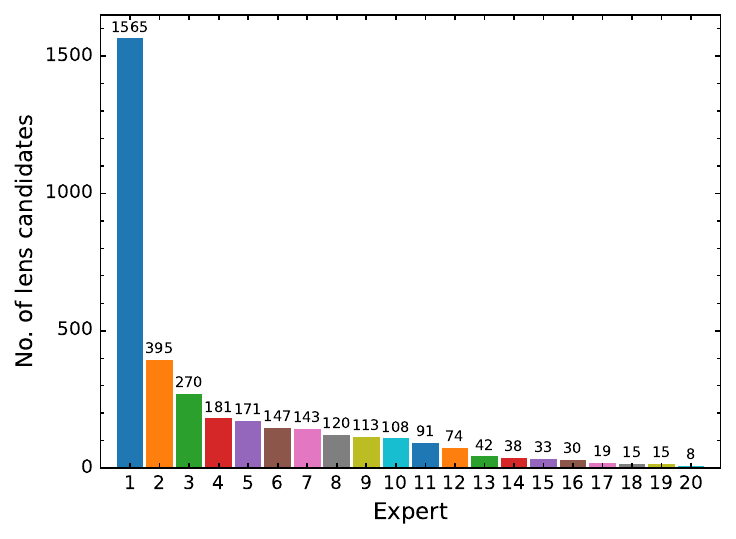}
\caption{Number of candidates selected in the first stage for reinspection by Team~1 using the mosaic tool.
The experts were anonymised, and the expert IDs correspond only to the first stage of the visual inspection.
}
\label{fig:team_1_stage_1_users}
\end{figure}

\begin{figure}[t!]
\centering
\includegraphics[width=0.5\textwidth]{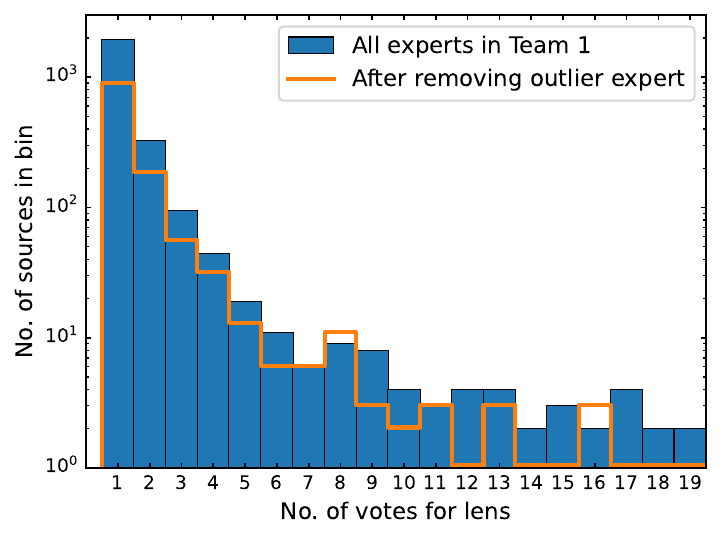}
\caption{Distribution of the number of positive votes for Team~1 before and after removing the outlier expert.
Most of the sources received only one vote, and no object was selected by all the experts.
}
\label{fig:team_1_stage_1_candidates}
\end{figure}

The first stage of the visual inspection was completed by 20 out of the 22 experts that registered for Team~1.
\Cref{fig:team_1_stage_1_users} presents the number of lens candidates selected by each expert.
We note that expert number 1 selected $15\,\sigma$ more sources for reinspection when compared to the mean and standard deviation of the other experts from Team~1, which selected $106\pm100$ sources for reinspection.
\Cref{fig:team_1_stage_1_candidates} presents the distribution of votes with and without expert number 1.
We observe that the number of sources selected only by a single expert drops by half when excluding the outlier expert.
Consequently, we removed the classifications from expert number 1.
This provided us with a generous selection for the second stage, in which we included every source chosen by any of the remaining experts, while also keeping the sample reasonably small.
The total number of sources selected for inspection in stage~2 was \num{1233}.

The second stage was completed by 17 out of the 22 experts.
\Cref{fig:team_1_stage_2_grades} presents the number of votes per grade per expert.
Similarly, \cref{table:results_stage_2} presents the average number of votes per grade for every stage that used the one-by-one sequential tool.
Both \cref{fig:team_1_stage_2_grades,table:results_stage_2} show that the experts of Team~1 were conservative when assigning the highest grade, A.
This trend, however, was not the case for grades B and C, for which multiple experts assigned hundreds of votes in the \num{1233} sample.
This indicates confusion among some experts for what constitutes a lens when the lensing features are not strikingly obvious.
However, the experts with a large number of positive classifications were still a minority.
Ultimately, the behaviour of all experts during the second stage was deemed acceptable and none of classifications were removed.

\subsection{Team 2}

\begin{table}[thbp!]
\centering
\caption{Average number of votes per grade per expert for the inspections using the one-by-one sequential tool. We use one standard deviation for the error range.}
\smallskip
\label{table:results_stage_2}
\smallskip
\begin{tabular}{lccc}
\hline
& & &\\[-9pt]
&  \multicolumn{1}{l}{First stage} & \multicolumn{2}{l}{Second stage} 
  \\ \hline
& & & \\[-9pt]
Grade & Team~2 & Team~1 & Team~2\\ \hline
& & &\\[-8pt]
A & $\phantom{4}4 \pm 4$ & $10 \pm 10$ & $10 \pm 10$\\
B & $\phantom{4}40 \pm 40$ & $\phantom{4}70 \pm 100$ & $40 \pm 40$\\
C & $\phantom{4}200 \pm 200$ & $200 \pm 300$ & $200 \pm 100$\\
X & $3800 \pm 200$ & $900 \pm 300$ & $500 \pm 200$\\
\hline
& & &\\[-9pt]
Total & $4031$ & $1233$ & $691$\\
\hline
\end{tabular}
\end{table}

\begin{figure*}[htbp!]
  \centering
\includegraphics[width=0.49\textwidth]{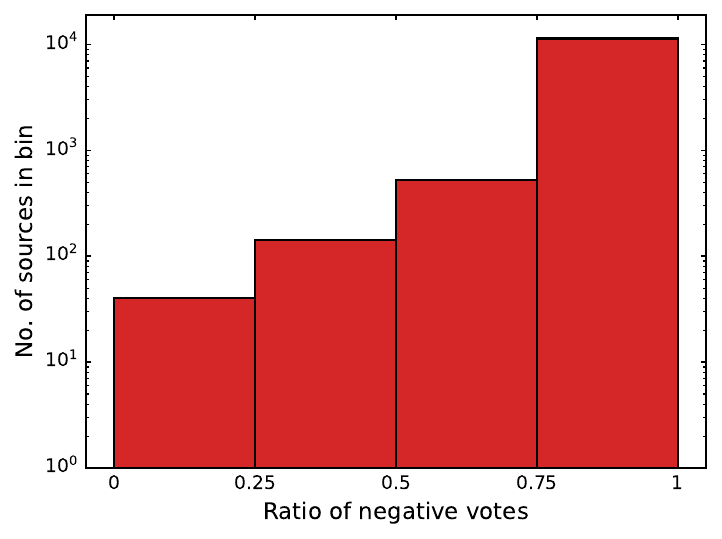}
\includegraphics[width=0.49\textwidth]{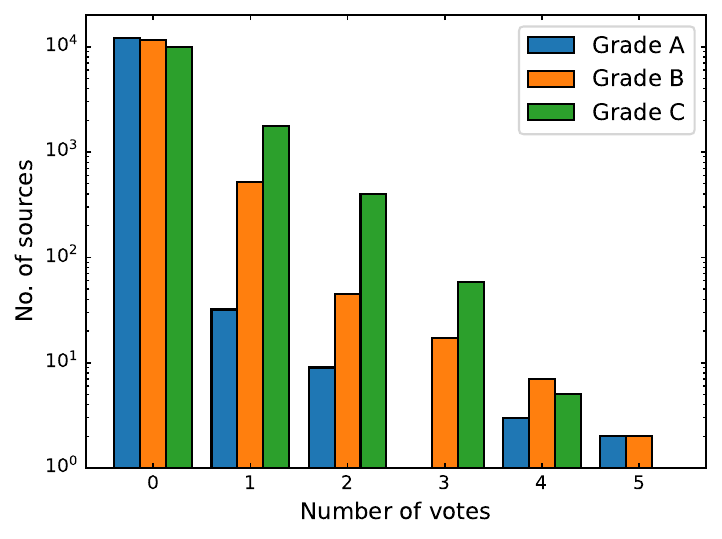}
\caption{Results from Team~2 during the first stage.
\emph{Left}: Histogram of the ratio of negative votes versus the total number of classifications for the source.
The sources with a ratio of negative votes lower than 0.5 were selected for stage~2.
\emph{Right}: Total number of sources that received a given number of votes for a given classification.
Most of the sources had zero votes for A, B, and C.
All sources with at least one vote for A or B were also selected for stage~2.
}
\label{fig:team_2_rung_1_selection_criterias}
\end{figure*}

The first round of visual inspection was completed by 16 out of 19 experts involved in Team~2.
The distribution among groups was as follows: five experts from both the first and second group, and another six from the third group.
\Cref{fig:team_2_stage_1_grades} and \cref{table:results_stage_2} summarise the behaviour of the experts during the inspection.
The grades A and B were given to only a small number of sources and the overall proportion of positive votes was low.
This is consistent with the task of removing the contaminants, which was the goal for stage~1.
The visual inspection results are presented in \cref{fig:team_2_rung_1_selection_criterias}.
We discarded all the sources with a negative majority vote, except those that got at least one vote for A or B.
This amounted to \num{691} sources for reinspection in stage~2.

The second stage of Team~2 was completed by 17 experts out of the 19 who registered.
The grading behaviour of the experts is presented in \cref{fig:team_2_stage_2_grades}.
A few of the experts gave a large number of positive grades, one grading positively 673 out of the 691.
Still, most of the experts reserved the grades A and B for the few best candidates, and thus, we did not remove any experts.
This is justified by the aggregation method introduced in \cref{sec:visual_inspections}, which relies on a majority vote to assign final grades.

\subsection{Time cost of the inspections}
\label{sec:comparison_of_the_teams}

It took the experts of Team~1 between $1\,{\rm h}\,15\,{\rm m}$ and $5\,{\rm h}\,15\,{\rm m}$ to inspect the whole parent sample using the mosaic tool, with the median time being $2\,{\rm h}\,45\,{\rm m}$.
By contrast, it took the experts from Team~2 between $1\,{\rm h}\,30\,{\rm m}$ to $3\,{\rm h}\,30\,{\rm m}$ with a median time of $2\,{\rm h}\,15\,{\rm m}$ to inspect one third of the parent sample, but using the one-by-one sequential tool.
The median time per source was $0.8\,{\rm s}$ and $2.0\,{\rm s}$ for Teams~1 and 2, respectively.

For the second stage, Team~1 experts took between $30\,{\rm m}$ and $3\,{\rm h}\,30\,{\rm m}$. The median time was $1\,{\rm h}\,30\,{\rm m}$, equivalent to $4.4\,{\rm s}$ per source.
Finally, Team~2 experts took between $15\,{\rm m}$ and $2\,{\rm h}$, and the median was $45\,{\rm m}$, which corresponds to $4.0\,{\rm s}$ per source.
Overall, the time taken per source during the second stage was consistent for both teams.
Given the filtering of the sample during the first stage, the time taken per source in the second stage corresponds more closely to the real time that a visual inspection would take on a preselected sample, for example the output of a convolutional neural network.

\section{Results}
\label{sec:results}
\subsection{The visual inspection candidates}
We computed three grades for each source: two individual grades from Team~1 and Team~2 plus a final joint grade.
The final grade of each team is determined by applying the scheme presented in \cref{sec:visual_inspections} to their respective set of votes.
The joint final grade is determined the same way but using the votes of both teams together in the computation,
with the exception of sources that were rejected by one of the teams during the first stage.
In that case, we added two negative votes to the classifications of the other team and calculated the final joint classification using the 19 votes.

\begin{figure}[htbp!]
  \centering
\includegraphics[width=0.50\textwidth]{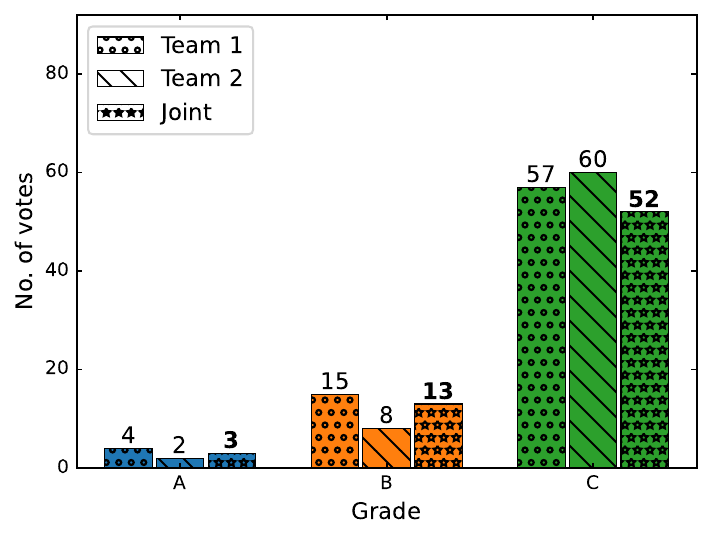}
\caption{Distribution of final grades for each team and for the joint classification.
}
\label{fig:barchart_final_grades}
\end{figure}

\begin{figure*}[htbp!]
\centering
\includegraphics[width=7.2in]{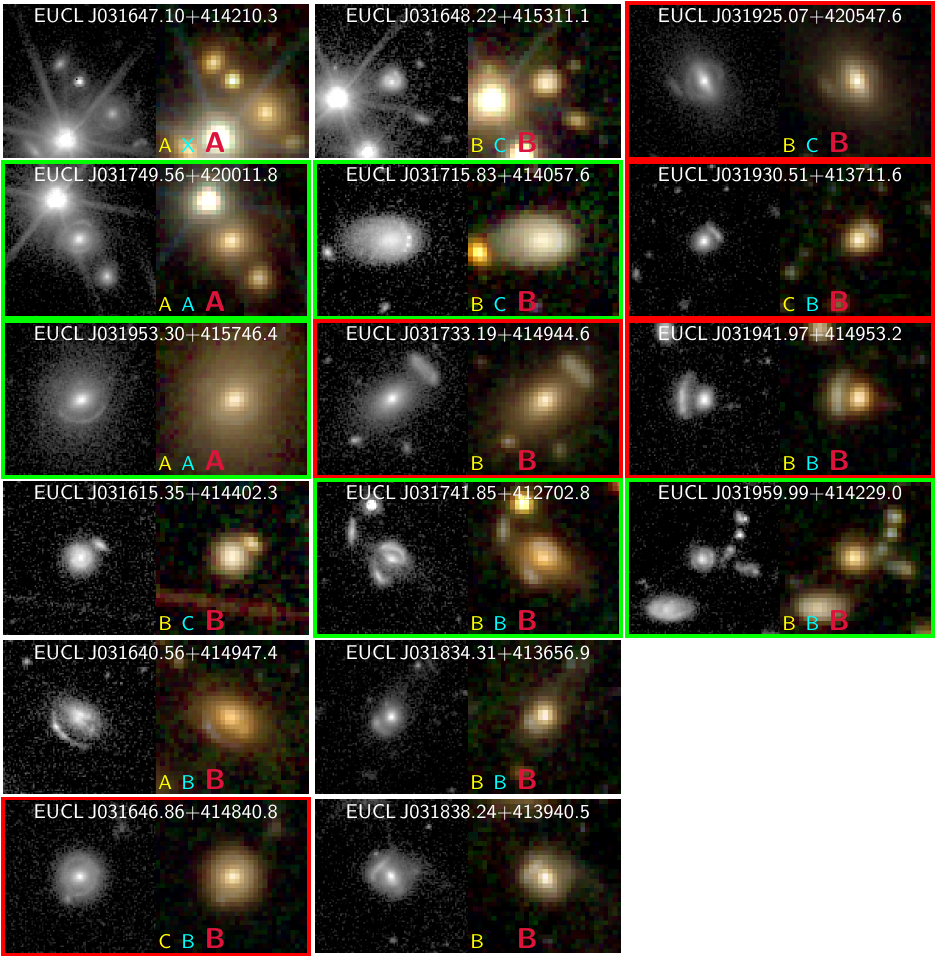}
\caption{Mosaic of the grade A and B lens candidates from the visual inspection.
For each candidate we show the high-resolution \IE band cutout in the left panel, and the lower-resolution \HE\YE\IE composite image in the right panel.
The final grades by Team~1, Team~2, and the final joint grade are shown in yellow, cyan, and red, respectively.
No grade is shown if one of the sources was rejected by a team during the first stage of the visual inspection.
The green borders highlight candidates with valid models, and red borders candidates rejected due to the modelling.
}
\label{fig:candidates_A_and_B}
\end{figure*}

Overall, we obtain three grade A, 13 grade B, and 52 grade C lens candidates from the final joint grades.
We present a summary of the number of candidates per team in \cref{fig:barchart_final_grades}.
Furthermore, we show the grade A and B candidates in \cref{fig:candidates_A_and_B}, along with their three final grades: the final grade from Team~1, the final grade from Team~2, and the final joint grade (in yellow, cyan, and red, respectively).
The grade C candidates, along with the sources selected by either team but rejected from the joint sample, are presented in a \href{https://zenodo.org/records/14946028}{Zenodo appendix}.
We note that Team~1 found almost double the amount of grade A and grade B candidates, but about the same number of grade C candidates as Team~2.
Moreover, all the candidates selected by Team~1 but rejected from the final sample were also rejected by Team~2 during the first stage.
By contrast, most of the sources selected by Team~2 but rejected from the final sample were also rejected by Team~1 during the second stage.
This suggests that using the one-by-one sequential tool to filter out the obvious non-lenses, as Team~2 did during the first stage, is in fact a more aggressive filter than using the mosaic tool to preselect candidates.

\subsection{Modelling of the lens candidates}
\label{sec:modelling}
We further assessed the validity of the 16 grade A and B candidates by modelling the lens galaxy light and mass distribution, and surface brightness distribution of the lensed source.
We did this using the \texttt{pronto} software \citep{Vegetti2009,Rybak2015a,Rybak2015b,Rizzo2018,Ritondale2019,Powell2021}. We used a singular isothermal ellipsoid (SIE) density profile to describe the mass distribution of the galaxy, a composite of three S\'ersic profiles for the lens galaxy light, and a pixelated, free-form reconstruction of the source surface brightness distribution. The parameters for the lens light and mass models are found through a non-linear optimisation using \texttt{MultiNest} \citep{Feroz2009}. For each set of light and mass parameters, the surface brightness distribution of the source is solved for linearly, up to some regularisation condition, which, in this case, penalises large gradients in the source. The strength of this regularisation is itself a non-linear parameter. The full model has 23 free parameters.

We used a circular mask centred on the lensing galaxy and large enough to enclose what are assumed to be lensed images.
We also used the positions of these assumed lensed images as an input to the optimisation scheme.
After selecting two, three, or four positions in the image plane, the non-linear optimiser will only accept models where these image plane positions have a root mean square separation in the source plane below some tolerance, in this case 1~arcsec.
This removes the need to fine-tune the initial modelling conditions, since only combinations of lens parameters that focus the source are accepted.
The large tolerance of 1~arcsec also prevents the subjective choice of image positions from placing strict restrictions on the model.

After the optimisation, we checked each model against three criteria to determine if its data are well described by a strong lensing model. These criteria are:
\begin{enumerate}
    \item  Is there an SIE critical curve that can enclose or exclude the right number of bright components in the image plane?
    \item Is the centroid of this critical curve consistent with that of the light profile?
    \item Is the reconstructed source surface brightness distribution consistent with a compact, focused object, inside a caustic?
\end{enumerate}
Given the small sample size, the evaluation of these criteria is done in a qualitative basis through the visual inspection of the models (see \cref{appendix:modelling-results}).
A systematic and quantitative assessment requires evaluating these criteria against realistic simulations where the ground-truth is known, which is beyond the scope of this work.

\begin{figure}
\includegraphics[width=1.0\columnwidth]{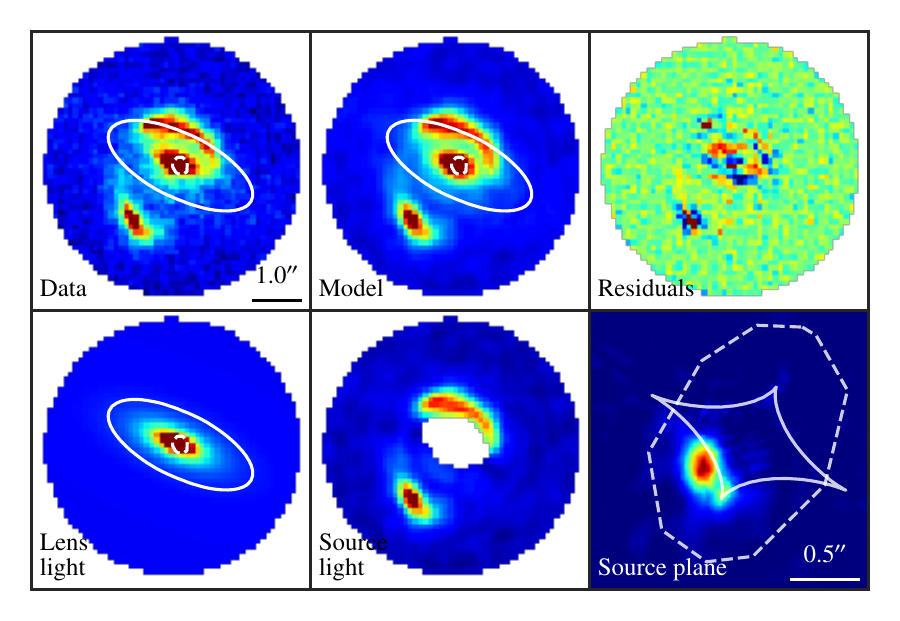}
\caption{\label{fig:valid-models} Modelling results for one of the five valid candidates, EUCL\,J031741.85+412702.8. The six panels show, from left to right, top to bottom: the VIS cutout data, masked to a circle around the object; the model lens light and lensed source light distributions, convolved with the point spread function model; the normalised residuals on a colour scale between $-3\,\sigma$ and $3\,\sigma$; the lens light only; the lensed source light only, with the brightest parts of the lens light masked; and the reconstructed source plane model. The solid and dashed white ellipses in the image plane are the tangential and radial critical curves, with the corresponding caustics shown in the source plane. The image plane frames all have the same scale, as indicated in the first frame.}
\end{figure}

Five of the 16 grade A and B candidates are found to have valid lens models according to the criteria above. Six candidates cannot be determined as lenses with the available data, and a final five candidates do not have valid lens models.
For the five valid candidates, we measure Einstein radii in the range \mbox{\ang{;;0.68}--\ang{;;1.24}}.
The result for one valid candidate is shown, as an example, in \cref{fig:valid-models}; the rest are shown in \cref{appendix:modelling-results}.
The results of the visual inspection and modelling are summarized in \cref{table:candidates}.

\begin{table*}[thbp!]
\centering
\caption{Grade A and B candidates with magnitude, modelling status, and Einstein radii when available.
The magnitudes correspond to the \texttt{MAG\_AUTO} key from the \ac{ERO} catalogues.}
\smallskip
\label{table:candidates}
\smallskip

\begin{tabular}{lllllll}
\hline
& & & & & &\\[-10pt]
Name & RA & Dec & \IE & Grade & Modelling & $\theta_\mathrm{E}$ [arcsec] \\\hline 
& & & & & &\\[-8.5pt]
EUCL\,J031715.83+414057.6 & 49.315974 & 41.682676 & $20.810 \pm 0.004$ & B & Valid &0.70 \\
EUCL\,J031741.85+412702.8 & 49.424405 & 41.450783 & $22.393 \pm 0.007$ & B & Valid &1.01 \\
EUCL\,J031749.56+420011.8 & 49.456529 & 42.003291 & $19.791 \pm 0.003$ & A & Valid &0.92 \\
EUCL\,J031953.30+415746.4 & 49.972097 & 41.962896 & $19.576 \pm 0.003$ & A & Valid &1.24 \\
EUCL\,J031959.99+414229.0 & 49.999999 & 41.708073 & $22.467 \pm 0.007$ & B & Valid &0.68 \\
\hline
& & & & & & \\[-9pt]
EUCL\,J031615.35+414402.3 & 49.063959 & 41.733984 & $21.651 \pm 0.005$ & B & Undetermined & \\
EUCL\,J031640.56+414947.4 & 49.169013 & 41.829837 & $21.902 \pm 0.007$ & B & Undetermined & \\
EUCL\,J031647.10+414210.3 & 49.196262 & 41.702880 & $21.193 \pm 0.003$ & A & Undetermined & \\
EUCL\,J031648.22+415311.1 & 49.200924 & 41.886438 & $21.171 \pm 0.005$ & B & Undetermined & \\
EUCL\,J031834.31+413656.9 & 49.642982 & 41.615827 & $21.280 \pm 0.002$ & B & Undetermined & \\
EUCL\,J031838.24+413940.5 & 49.659347 & 41.661254 & $20.905 \pm 0.002$ & B & Undetermined & \\
\hline
& & & & & & \\[-9pt]
EUCL\,J031646.86+414840.8 & 49.195286 & 41.811343 & $20.837 \pm 0.002$ & B & Not valid & \\
EUCL\,J031733.19+414944.6 & 49.388333 & 41.829079 & $20.666 \pm 0.002$ & B & Not valid & \\
EUCL\,J031925.07+420547.6 & 49.854480 & 42.096580 & $19.461 \pm 0.001$ & B & Not valid & \\
EUCL\,J031930.51+413711.6 & 49.877131 & 41.619889 & $22.147 \pm 0.004$ & B & Not valid & \\
EUCL\,J031941.97+414953.2 & 49.924888 & 41.831457 & $22.014 \pm 0.003$ & B & Not valid & \\
\hline
\end{tabular}
\end{table*}

\section{Discussion}
\label{sec:discussion}

\subsection{The lens candidates in the literature}
We crossmatched the sky coordinates of the 68 lens candidates against the Strong Lens Database (SLED; Vernardos et al., in prep.), a database of gravitational lenses, lens candidates, and known contaminants, encompassing more than \mbox{\num{20000}} entries.
We found no matches between our sample and the database.
Furthermore, the database indicates that there have been no lens searches in the area, since the closest match is more than $\ang{6}$~away. 
This is expected given that the Perseus cluster is near the Galactic plane, and thus, is not generally targeted by large-scale optical galaxy surveys.

Nonetheless, \cite{2024AJ....167..264L} have performed a lens search in some of the \ac{ERO} fields, including the Perseus cluster, in order to test the lens detection algorithm to be used by the \ac{CSST} team.
For that purpose, they used the high-resolution media images published by ESA.\footnote{Available here: \url{https://www.esa.int/Science_Exploration/Space_Science/Euclid/Euclid_s_first_images_the_dazzling_edge_of_darkness}}
They found four lens candidates in the Perseus cluster and reported the pixel coordinates in the TIFF image.
Three of those candidates are included in our parent sample and were selected by Team~1 for reinspection in the second stage.
However, all of them were rejected in the second stage.
A smaller committee of experts inspected the fourth candidate and ultimately rejected it.
Thus, we believe that none of the \cite{2024AJ....167..264L} candidates in the Perseus cluster are lenses.
The lack of real lenses in their sample is likely a consequence of using the lower-resolution TIFF images and the fact that their model was not trained to detect lenses in \Euclid data.
Therefore, the false positives in the Perseus cluster do not necessarily reflect on the potential performance of the algorithm in \ac{CSST} data.

\subsection{The expected prevalence of ELSE lenses}

\citet[C15 hereafter]{Collett2015} forecasts that $\num{15000}\,{\rm deg^2}$ of \Euclid imaging should contain \mbox{\num{170000}} strong gravitational lenses.
Naively scaling down to the $0.7\,{\rm deg^2}$ Perseus field gives an expectation of eight galaxy-galaxy lenses, of which six should have \IE < 23. We have three grade A and 13 grade B candidates, of which five have a valid lens model, and six are indeterminate.
Despite the small number statistics, it is clear that our results and the C15 forecasts are broadly consistent.
As well as the Poisson noise, our limited understanding of the discovery selection function, and the lack of spectroscopic redshift confirmation of our candidates makes it impossible to precisely compare our absolute number of lenses with the forecast population.

\begin{figure}[t!]
\centering
\includegraphics[width=0.5\textwidth]{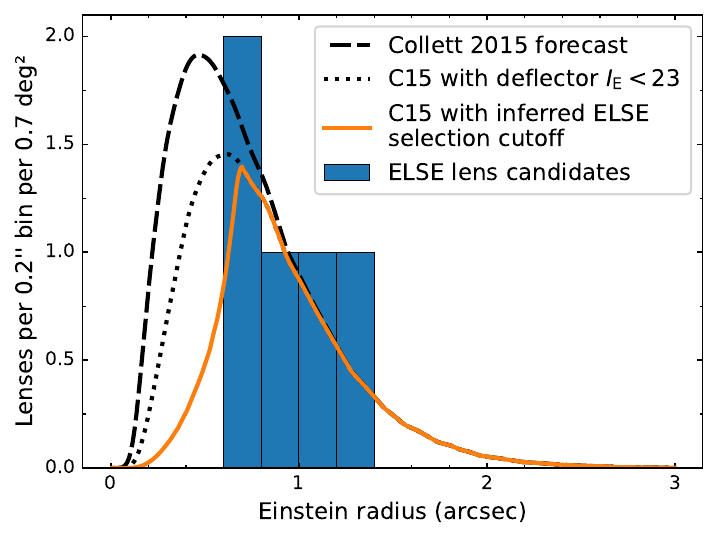}
\caption{Comparison of the number of lenses and their Einstein radius distribution of successfully modelled ELSE candidates (blue histogram), and the forecasts of \citet{Collett2015} rescaled to $0.7\,{\rm deg^2}$ (dotted black line). The dotted black line is a prediction, not a fit. The orange `cutoff' line is a modification of the \citet{Collett2015} forecasts to account for the selection function arising from our methodology missing small-Einstein-radius lenses. 
}
\label{fig:forecast}
\end{figure}

The median Einstein radius forecast in C15 is \mbox{\ang{;;0.65}}, whereas the smallest Einstein radius of the five lenses with valid models in \cref{sec:modelling} is \mbox{\ang{;;0.68}}. This tension is alleviated somewhat by accounting for the fact we only look at lenses with \IE < 23: by excluding the faintest lenses we prefer higher-mass or lower-redshift lenses, both of which result in larger Einstein radii. After applying this cut, the forecast median Einstein radius increases to \mbox{\ang{;;0.75}}.
\Cref{fig:forecast} shows the expected number of lenses in $0.7\,{\rm deg^2}$ of \Euclid data, their Einstein radius distribution, and the Einstein radius distribution of successfully modelled ELSE candidates.
The C15 forecasts assumed that the lens galaxy light would be subtracted, which is very helpful for finding small-Einstein-radius lenses: we have not done this.
In future work, we will calibrate this selection function using simulations (Rojas et al., in prep.); in this study we used a toy model to estimate it.
We assumed that the shape of the underlying Einstein radius population follows the \Euclid forecasts of C15, and that the visual inspection introduced a selection function in the form of a step function, we detect every lens above a threshold Einstein radius ($\theta_\mathrm{min}$) and none below this threshold.
Using Bayes' theorem, we infer $\prior{\theta_\mathrm{min}}$ given that we discovered five lenses and the lowest Einstein radius is in the range \mbox{\ang{;;0.65}}--\mbox{\ang{;;0.70}}.

We used a Monte Carlo simulation to draw many realisations of five Einstein radii but varied the discovery threshold, assuming a uniform prior on the discovery threshold, $\theta_\mathrm{min}$.
We find that the threshold is $0.57_{-0.26}^{+0.08}$~arcsec at 68\% confidence. This corresponds to a forecast total population of $\num{93000}_{-\num{13000}}^{+\num{34000}}$~lenses in the full \Euclid dataset that are discoverable with visual inspection without lens light subtraction and with lens \IE < 23. 
Moreover, scaling down the area to $0.7\,{\rm deg}^2$ the forecast predicts approximately five lenses, coinciding with the five lens candidates with good models.

\subsection{The prevalence of ELSE lenses}
\label{sec:hypergeometric_prevalence}

We extrapolated the five lens candidates with valid lens models found in the ERO Perseus field to
the entire \ac{EWS}.
For this, we made three key assumptions:
\begin{enumerate}
    \item We have found all the discoverable lenses with $\IE<23$ within the ERO Perseus field.    \item The average number density of sources with $\IE<23$ in the \ac{EWS} is the same as in the ERO Perseus field: ${\sim}\,\num{17000}$ extended sources per ${\rm deg}^{2}$.\footnote{About $5\%$ of sources are cluster members and they occlude about $5\%$ of the sky. In terms of number of sources, the effect of the cluster balances out.}
    \item The likelihood that a given source within the \ac{EWS} exhibits lensing features mirrors that of the sources within the parent sample.
\end{enumerate}
We defined a lens as discoverable if it can be identified via expert visual inspection and a valid lens model can be found for it.
This assumption means that, in fact, we are mostly estimating the prevalence of larger Einstein radius lenses, which are easier to model and experts will identify more easily.
However, small-Einstein-radius lenses can still be identified if the source is bright and the geometry of the lensed images is particularly evident; for example, the mock lens in the bottom right panel of \cref{fig:all_mocks}, which has an Einstein radius of \mbox{\ang{;;0.49}}, but also a typical quad geometry, was still graded A by the experts.
Furthermore, given the increased depth of the \ac{ERO} Perseus field data compared to the \ac{EWS} (0.75 magnitudes), we simulated cutouts for the 16 lens candidates with the typical S/N of the \ac{EWS}.
We observe that the lensing features are still visible and clearly identifiable in the shallower cutouts.
Consequently, we assumed that they would had been identified in the shallower \ac{EWS}.

\begin{figure}[t!]
  \centering
\includegraphics[width=0.50\textwidth]{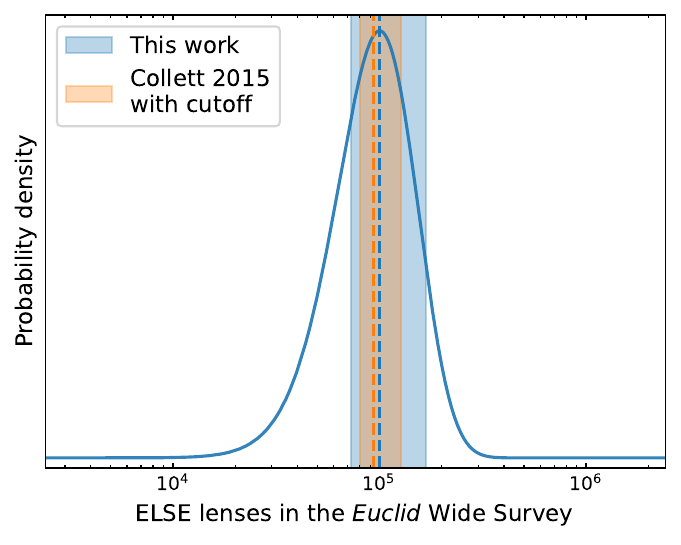}
\caption{Posterior probability density distribution of our estimation of the number of ELSE lenses in the \ac{EWS}.
The blue-shaded region marks the 68\% confidence interval of our estimation: $\num{100000}_{-\num{30000}}^{+\num{70000}}$ ELSE lenses in the \ac{EWS}.
The orange-shaded region marks the 68\% confidence interval of the \citet{Collett2015} forecast after applying our selection function:
$\num{93000}_{-\num{13000}}^{+\num{34000}}$.
}
\label{fig:else_prevalence_posterior}
\end{figure}

Under the key assumptions above, the prevalence of ELSE lenses in the \ac{EWS} can be estimated as follows. The likelihood $\prior{k}$ of finding $k$ lenses among $n$ trialled sources is given by the hypergeometric distribution
\begin{equation}
    \prob{k}{K}=\frac{\binom{K}{k}\binom{N-K}{n-k}}{\binom{N}{n}},
\end{equation}
where $K$ and $N$ are the total numbers of lenses and sources in the \ac{EWS}.
Using the second assumption, we estimate the total number of extended sources with $\IE<23$ in the $\num{14000}\,{\rm deg}^2$ of the \ac{EWS} to be $N=\num{242000000}$.
The number of trialled sources was $n=\num{12086}$, and the number of discoverable lenses was $k=5$.
We could then compute the posterior probability, $\prior{K}$, using Bayes' theorem (i.e. $\prob{K}{k}\propto \prob{k}{K}\prior{K}$).
We assumed that the prior probability of $K$, $\prior{K}$, is such that the ratio $K/N$ is distributed uniformly in log-space between $10^{-10}$ and unity.
Overall, we expect $\num{100000}_{\num{-30000}}^{+\num{70000}}$ ELSE-type lenses in the entire \ac{EWS}.
The posterior distribution $\prior{K}$ is presented in \cref{fig:else_prevalence_posterior}, with the 68\% confidence interval highlighted in light blue.
\Cref{fig:else_prevalence_posterior} also shows in orange the forecast of C15 as described in the previous section, which lies inside the confidence range of our estimate, showing a good agreement between the forecast and our estimate.

\section{Conclusions}
\label{sec:conclusions}

In this work we have investigated the performance of the \Euclid telescope for detecting galaxy-scale gravitational lenses.
We employed 41 experts to carry out a blind visual search for lenses in the \Euclid \ac{ERO} data of the Perseus cluster.
The search yielded 3 grade A and 13 grade B lens candidates.

We modelled the 16 candidates to test whether the observed VIS images are consistent with a single background galaxy lensed by a simple, plausible lens.
We have obtained convincing models for five of the candidates.
Modelling for six of the candidates produced inconclusive results, whilst five of the visual inspection candidates are definitively non-lenses according to our modelling. 

We extrapolated the five candidates with valid lens models to the entire \ac{EWS} and estimated the number of lenses in it that are discoverable with visual inspection and can be confirmed through modelling of the VIS data.
This extrapolation yields $\num{100000}^{+\num{70000}}_{-\num{30000}}$ lenses in the full \ac{EWS}.
This is broadly consistent with the \mbox{\num{170000}} forecast by C15.

Even though our magnitude cut of $\IE<23$ effectively removes many of the small-Einstein-radius objects, the distribution of Einstein radii in our modelled sample is still on the higher side when compared to C15, indicating that we are either unable to identify small-Einstein-radius systems in the visual inspections or unable to model them under our simplifying assumptions.
Assuming a step function cutoff in Einstein radius, we inferred the cutoff point to be $0.57_{-0.26}^{+0.08}\,\rm{arcsec}$; below this, our methodology is likely missing most of the lens candidates.
Convolving our inferred cutoff with the \citet{Collett2015} \Euclid population, we now predict that $\num{93000}_{-\num{13000}}^{+\num{34000}}$ galaxy-scale lenses will be detected in the whole \ac{EWS} (assuming that the same visual inspection discovery and modelling methodology are performed on the entire dataset).
Down-scaling to the $0.7\,{\rm deg}^2$ visually inspected in this work gives approximately five lenses, which is in perfect accordance with the five lens candidates with valid models.

This sample represents the first gravitational lenses reported in this patch of the sky, and some of the first lenses discovered with \Euclid data.
There is tentative evidence that we are missing small-Einstein-radius lenses, so future work will be necessary to develop identification techniques targeting small-Einstein-radius lens candidates.
There are also substantial challenges in how we scale up from this field to the full dataset: blind visual inspection of $\num{14000}\,\rm{deg}^2$ is implausible.
Neural networks and citizen science are both likely to be needed to reduce the sample to a manageable size. 

Even without spectroscopic confirmation of our candidates, our simple lens modelling provides compelling evidence that at least five of them are indeed strong gravitational lenses. These results are hugely promising for the future of strong lensing science with \Euclid. Forecasts and early data both now point to the discovery of $10^5$ or more gravitational lenses in the full \Euclid dataset.

\section*{Data availability}
Supplementary materials showing the lens candidates discovered in the search but not included in the main sample
are available on Zenodo, at \href{https://zenodo.org/records/14946028}{https://zenodo.org/records/14946028}.

\begin{acknowledgements}
\AckERO
\AckEC
J.~A.~A.~B., B.~C., and F.~C. acknowledge support from the Swiss National Science Foundation (SNSF).
C.~O’R thanks the Max Planck Society for support through a Max Planck Lise Meitner Group.
C.~T. and V.~B. acknowledge the INAF grant 2022 LEMON.
T.~E.~C. is funded by a Royal Society University Research Fellowship. 
S.~H.~S. thanks the Max Planck Society for support through the Max Planck Fellowship. 
This research is supported in part by the Excellence Cluster ORIGINS which is funded by the Deutsche Forschungsgemeinschaft (DFG, German Research Foundation) under Germany’s Excellence Strategy -- EXC-2094 -- 390783311.
This project has received funding from the European Research Council (ERC)
under the European Union’s Horizon 2020 research and innovation
programme (LensEra: grant agreement No 945536 and LENSNOVA: grant agreement No 771776).
\end{acknowledgements}

\bibliographystyle{aa}
\bibliography{bibliography, Euclid_after_publication_02_05_2025 ,vegettietal}

\begin{appendix}

\section{Expert classifications in visual inspections}
We show in this appendix the detailed classifications given by the experts when using the one-by-one sequential tool.
\begin{figure}[htbp!]
\centering
\includegraphics[width=0.5\textwidth]{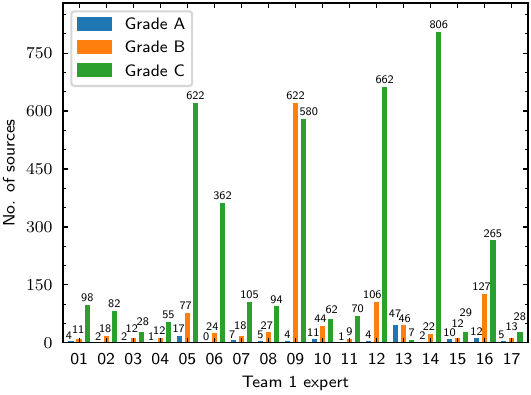}
\caption{Classifications given by Team~1 during the second stage of the visual inspection.
Each expert inspected \num{1221} sources and assigned the grades introduced in \cref{sec:visualization_tools}.
Negative grades (X) were not included, for the sake of simplicity.
The expert IDs correspond only to the second stage.
}
\label{fig:team_1_stage_2_grades}
\end{figure}

\begin{figure}[htbp!]
\centering
\includegraphics[width=0.5\textwidth]{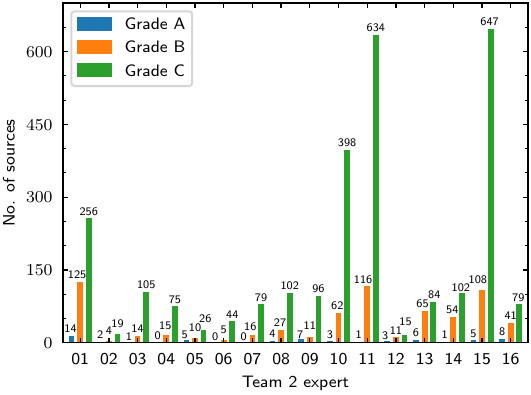}
\caption{Classifications given by Team~2 during the first stage of the visual inspection.
Each expert inspected about $4030$ sources and assigned the grades introduced in \cref{sec:visualization_tools}.
Negative grades (X) were not included, for the sake of simplicity.
The expert IDs correspond only to the first stage.
}
\label{fig:team_2_stage_1_grades}
\end{figure}

\begin{figure}[htbp!]
\centering
\includegraphics[width=0.5\textwidth]{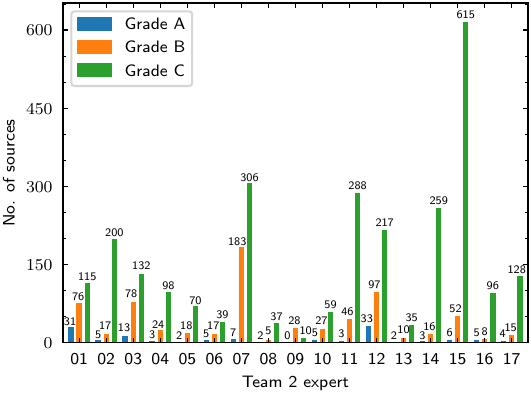}
\caption{Classifications given by Team~2 during the second stage of the visual inspection.
Each expert inspected \num{691} sources and assigned the grades introduced in \cref{sec:visualization_tools}.
Negative grades (X) were not included, for the sake of simplicity.
The expert numbers correspond only to the second stage.
}
\label{fig:team_2_stage_2_grades}
\end{figure}

\clearpage
\onecolumn
\section{Modelling results}
\label{appendix:modelling-results}
\begin{figure*}[htbp!]
    \centering
    \includegraphics[width=1.0\textwidth]{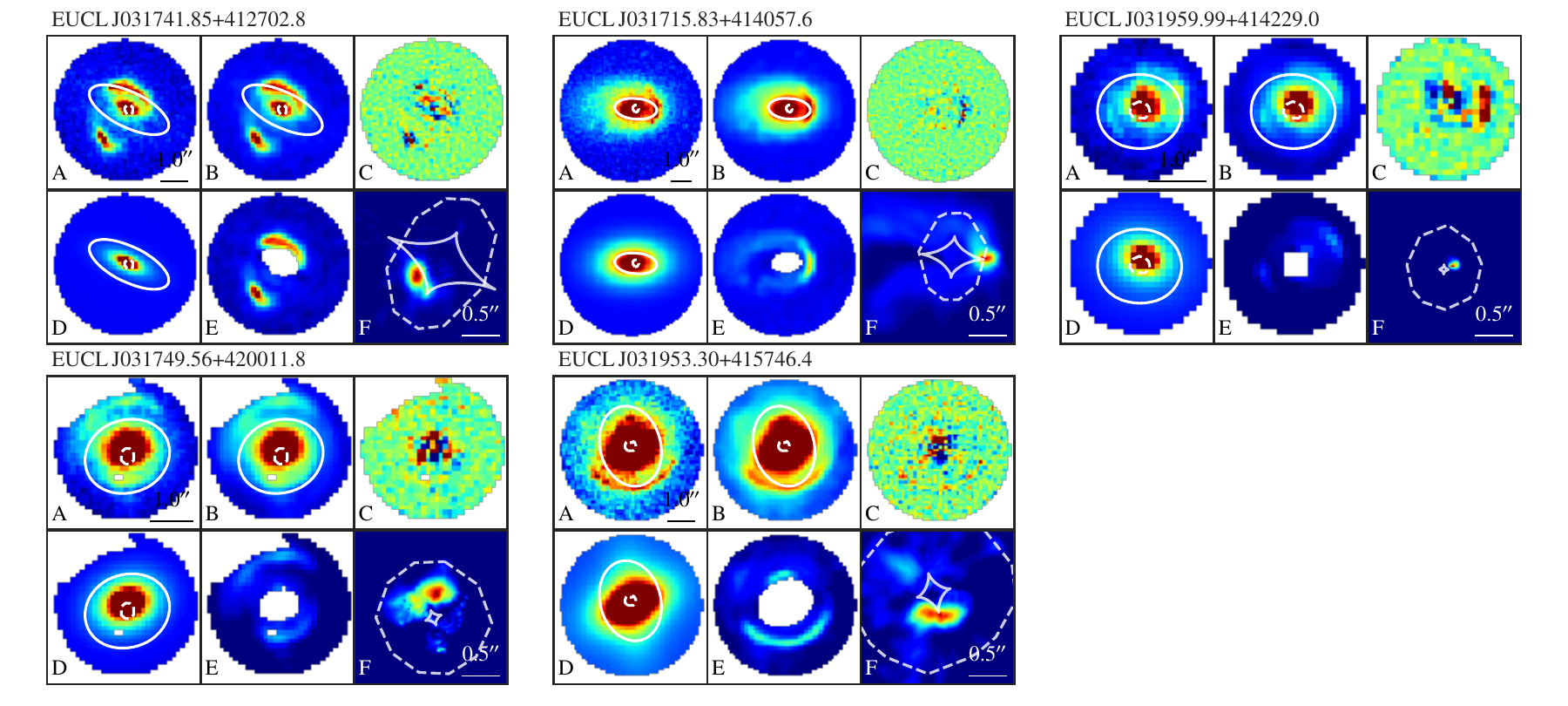}
    \caption{Modelling results for all five of the candidates that have valid lens models.
Panels (A): VIS cutout data, masked to a circle around the object.\ Panels (B): Model lens light and lensed source light distributions, convolved with the point spread function model. Panels (C): Normalised residuals on a colour scale between $-3\,\sigma$ and $3\,\sigma$. Panels (D): Lens light only. Panels (E): Lensed source light only, with the brightest parts of the lens light masked. Panels (F): Reconstructed source plane model. The solid and dashed white ellipses in the image planes are the tangential and radial critical curves, with the corresponding caustics shown in the source plane. The image plane frames all have the same scale, which is indicated in the first frame.}
    \label{fig:modelling_results_valid}
\end{figure*}

\begin{figure}[htbp!]
    \centering
    \includegraphics[width=1.0\textwidth]{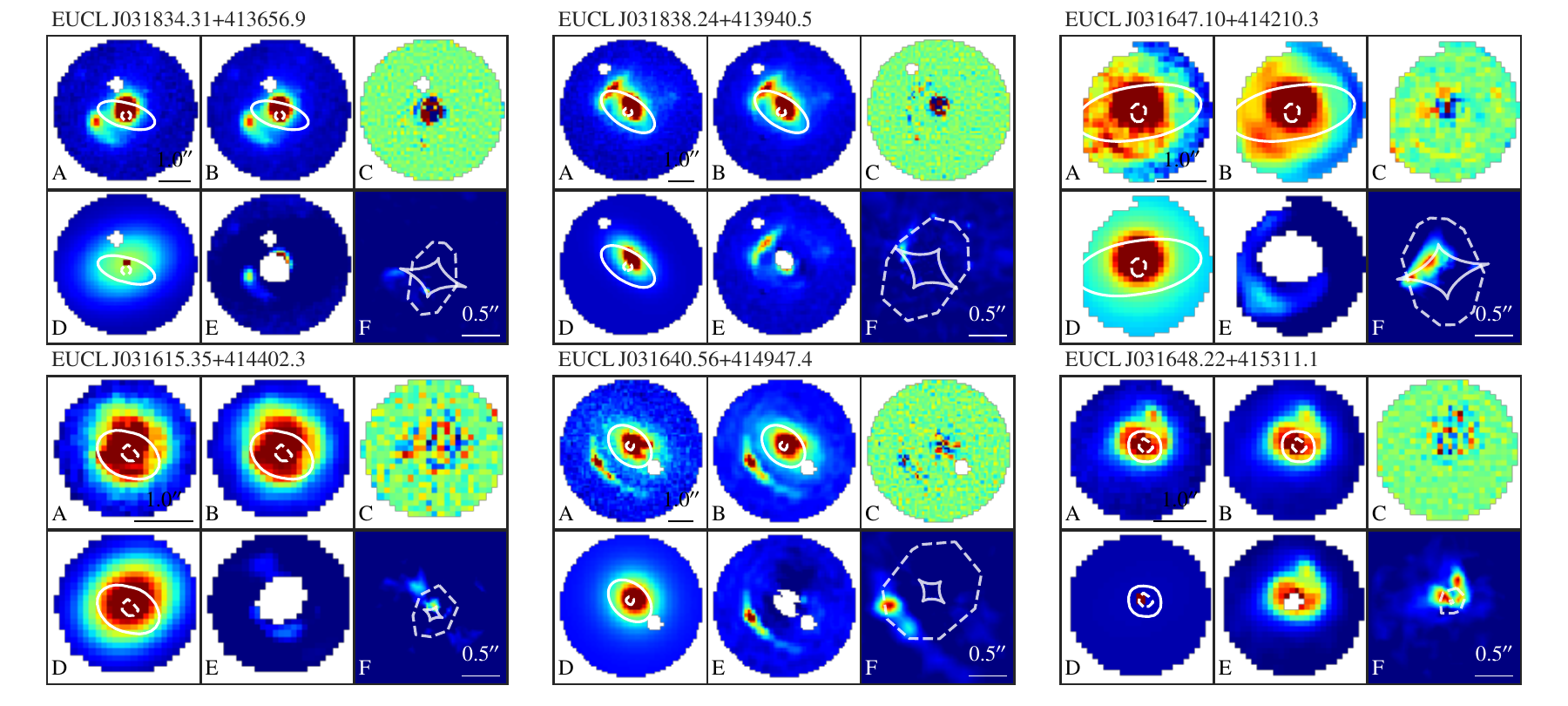}
    \caption{Modelling results for the six candidates for which we cannot determine their lensing status by modelling the current data.
    Mosaics follow the same layout as \cref{fig:modelling_results_valid}.}
    \label{fig:modelling_results_uncertain}
\end{figure}

\begin{figure}[htbp!]
    \centering
    \includegraphics[width=1.0\textwidth]{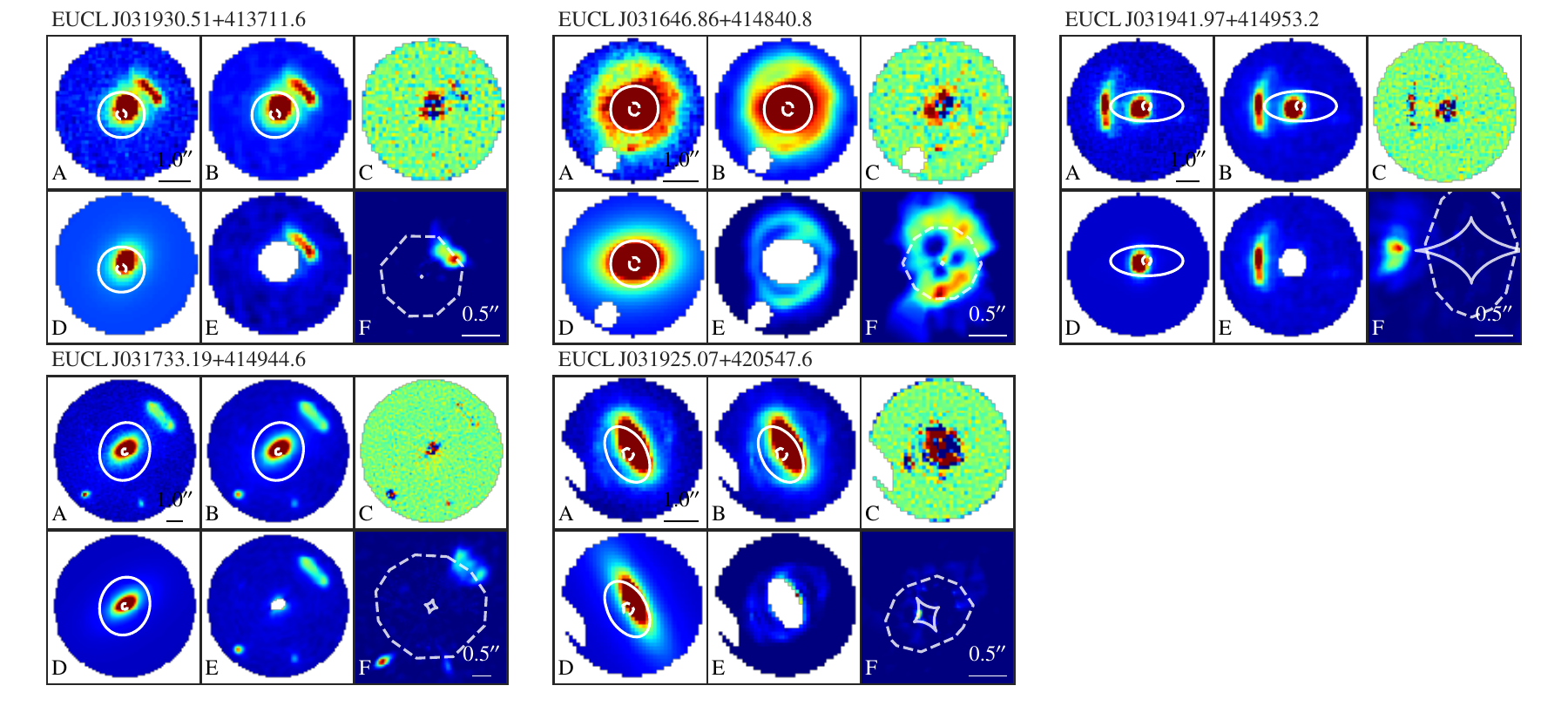}
    \caption{Modelling results for the five candidates that do not have valid lens models.
    Mosaics follow the same layout as \cref{fig:modelling_results_valid}.}
    \label{fig:modelling_results_not_valid}
\end{figure}

The results of modelling all 16 grade A and B candidates are shown in \cref{fig:modelling_results_valid,fig:modelling_results_uncertain,fig:modelling_results_not_valid}.

\newpage

\end{appendix}
\label{LastPage}
\end{document}